\documentclass[11pt,twocolumn]{article}
\usepackage{lmodern}   % Improved font rendering
\usepackage{titling}   % Custom title modifications
\usepackage{authblk}   % Better author formatting
\usepackage{fancyhdr}  % Header and footer customization
\usepackage{footmisc}

\pretitle{\begin{center}\LARGE\bfseries} % Large bold title
\posttitle{\par\end{center}\vskip0.5em}
\makeatletter
\renewcommand{\@fnsymbol}[1]{\ifcase#1\or \dag\or \ddag\or *\or §\or ¶\or **\or ††\or ‡‡\else \@ctrerr\fi}
\makeatother
\newcommand{\keywords}[1]{%
    \vskip1em
    \noindent\textbf{Keywords:} #1
    \vskip1em
}
\usepackage{titlesec}
\titleformat{\section}[hang]   % 'hang' aligns the title to the left
{\normalfont\Large\bfseries}   % Use normal font, smaller size, and bold
{\thesection}{1em}{}           % Numbering and space between number and title

\titleformat{\subsection}[hang] % 'hang' aligns the title to the left
{\normalfont\large\bfseries}    % Use normal font, smaller than small, bold
{\thesubsection}{1em}{}         % Numbering and space between number and title

\titleformat{\subsubsection}[runin] % Command shape
{\normalfont\bfseries} % Format
{\thesubsubsection}{1em}{\addperiod} % Label
\newcommand{\addperiod}[1]{#1.}

\usepackage{caption}  % For customizing captions
\usepackage{amsmath,amssymb,amsfonts,amsthm,bm}
\makeatletter
\def\tagform@#1{\maketag@@@{[\ignorespaces#1\unskip\@@italiccorr]}}
\makeatother

\usepackage{graphicx,xcolor,hhline}
\usepackage[separate-uncertainty=true,range-phrase=--,range-units=single]{siunitx}
\usepackage[colorlinks,linkcolor={blue},citecolor={blue},urlcolor={blue}]{hyperref}
\usepackage{fancyhdr}
\usepackage[utf8]{inputenc}
\usepackage[hmargin=1.2cm,vmargin=2cm]{geometry}
\usepackage{orcidlink}

\usepackage[backend=biber,style=science,citestyle=science,sorting=none]{biblatex}
\newcommand{\avg}[1]{\langle#1\rangle}

\newcommand{\Fe}{F_{\rm e}}
\newcommand{\kbt}{k_{\rm{B}}T}
\newcommand{\Wa}{W_{\rm a}}
\newcommand{\We}{W_{\rm e }}

\newcommand{\SIAppendix}{Supplemental Information}
\newcommand{\fgFitting}{\ref{fig:time-series}} % All fitted time series
\newcommand{\fgMap}{\ref{fig:map}} % Map of the thermodynamic regimes in all four experimental cases
\newcommand{\fgFlows}{\ref{fig:flows}} % Map of flows
\newcommand{\fgQ}{\ref{fig:q}} % Heat maps of energy flows for all four experimental cases
\newcommand{\fgWa}{\ref{fig:wa}} % Active work in all four cases
\newcommand{\fgWe}{\ref{fig:we}} % External work in all four cases
\newcommand{\fgIn}{\ref{fig:in}} % Efficiency in
\newcommand{\fgSNR}{\ref{fig:snr}} % SNR
\newcommand{\fgLast}{\ref{fig:last}} % Case 3 and 4 near Hopf bifurcation
\newcommand{\eqHvdf}{\ref{eq:hvdf}}
\newcommand{\tbScaleOne}{\ref{tab:scaling1}} % Scaling by lambda one
\newcommand{\tbScaleTwo}{\ref{tab:scaling2}} % Scaling by lambda two

\usepackage{booktabs}
\usepackage{empheq}
\newcommand{\Ka}{K_{\rm{a}}}
\newcommand{\Xa}{X_{\rm{a}}}
\newcommand{\Xe}{x}
\newcommand{\Xo}{X_{\rm{off}}}
\newcommand{\Xae}{Y}

\newcommand{\ga}{\gamma_{\rm{a}}}
\newcommand{\Kg}{K_{\rm{GS}}}
\newcommand{\Ks}{K_{\rm{SP}}}
\newcommand{\Fm}{F_{\rm{max}}}
\newcommand{\kBT}{k_{\rm{B}}T}

\newcommand\Eq[1]{Eq.~[\ref{#1}]}
\newcommand\Tab[1]{Table~\ref{#1}}
\begin{document}

% Document Content
\title{Active energy harvesting and work transduction by hair-cell bundles in bullfrog's inner ear}

\author[1,{\orcidlink{0009-0004-4197-313X}}]{Yanathip Thipmaungprom}
\author[2,3,{\orcidlink{0009-0002-1292-2304}}]{Laila Saliekh}
\author[4,{\orcidlink{0000-0003-0237-2660}}]{Rodrigo Alonso}
\author[2,\dag,{\orcidlink{0000-0001-7196-8404}}]{\'{E}dgar Rold\'{a}n}
\author[1,\ddag,{\orcidlink{0000-0003-3355-4336}}]{Florian Berger}
\author[5,*,{\orcidlink{0000-0002-8896-8109}}]{Roman Belousov}

\affil[1]{Utrecht University, Utrecht 3584, The Netherlands}
\affil[2]{ICTP---The Abdus Salam International Centre for Theoretical Physics, Trieste 34151, Italy}
\affil[3]{University of Edinburgh, Edinburgh EH8 9YL, UK}
\affil[4]{The Rockefeller University, New York 10065, USA}
\affil[5]{EMBL---European Molecular Biology Laboratory, Heidelberg 69117, Germany}

\renewcommand{\abstractname}{}  % Remove the word "Abstract"
\date{\small\today} % Removes default date placements

\twocolumn[
    \begin{@twocolumnfalse} % Prevent title from splitting across columns
        \maketitle
        \begin{abstract}
        Hair cells actively drive oscillations of their mechanosensitive organelles---the hair bundles that enable hearing and balance sensing in vertebrates. Why and how some hair cells expend energy by sustaining this oscillatory motion in order to fulfill their function as signal sensors and others---as amplifiers, remains unknown. We develop a stochastic thermodynamic theory to describe energy flows in a periodically-driven hair bundle. Our analysis of thermodynamic fluxes associated with hair bundles' motion and external sinusoidal stimulus reveals that these organelles operate as thermodynamic work-to-work machines under different operational modes. One operational mode transduces the signal's power into the cell, whereas another allows the external stimulus to harvest the energy supplied by the cell. These two regimes might represent thermodynamic signatures of signal sensing and amplification respectively. In addition to  work transduction and energy harvesting, our model also substantiates the capability of hair-cell bundles to operate as heaters and, at the expense of  external driving, as active feedback refrigerators. We quantify the performance and robustness of the work-to-work conversion by hair bundles, whose efficiency in some conditions exceeds 80\,\% of the applied power.
        \end{abstract}
        \vskip1em % Add spacing before two-column starts
    \end{@twocolumnfalse}
]

\renewcommand{\thefootnote}{\dag}
\footnotetext[1]{Author email: \href{mailto:edgar@ictp.it}{edgar@ictp.it}}
\renewcommand{\thefootnote}{\ddag}
\footnotetext[2]{Author email: \href{mailto:f.m.berger@uu.nl}{f.m.berger@uu.nl}}
\renewcommand{\thefootnote}{*}
\footnotetext[3]{Corresponding author: \href{mailto:roman.belousov@embl.de}{roman.belousov@embl.de}}
\renewcommand{\thefootnote}{\arabic{footnote}} % Reset footnote numbering

\begin{figure}[t!]%[tbhp]
\centering
\includegraphics[width=\linewidth]{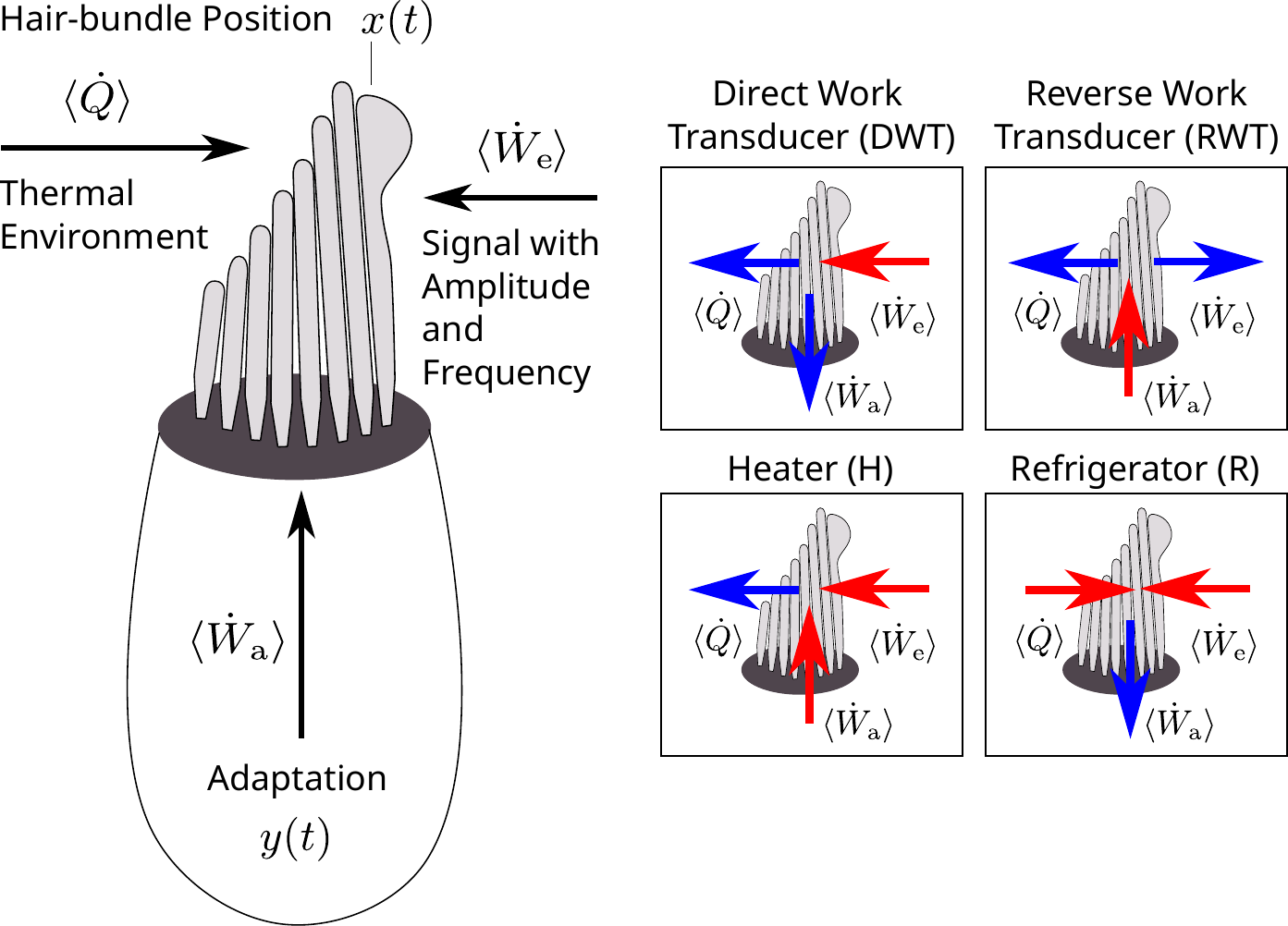}
\caption{Left: Schematic of a hair cell with its hair bundle on top. Energy is exchanged between the thermal environment, the hair bundle characterized by the position $x(t)$ of its tallest cilium---the kinocilium,---and two agents: an internal active adaptation described by the hidden variable $y(t)$, and an external signal of a certain amplitude and frequency. The average rates of energy exchange between the hair bundle and these three agents represent flows in the system: heat $\avg{\dot{Q}}$ exchanged with the environment, active power $\avg{\dot{W}_{\rm a}}$ applied by the internal adaptation, and external power $\avg{\dot{W}_{\rm e}}$ supplied by the signal. Black arrows indicate the positive sense of individual energy flows. Insets on the right illustrate four thermodynamic regimes, in which the hair bundle operates with blue and red arrows indicating the negative and positive sense of energy flows, respectively.}
\label{fig:intro}
\end{figure}

Our sense of hearing and balance is facilitated by hair bundles---mechanosensory organelles on the apical side of the namesake receptor cells of the inner ear (Fig.~\ref{fig:intro}). It is generally accepted that hearing relies on an active process, responsible for four cardinal features: amplification, frequency selectivity, compressive nonlinearity, and spontaneous otoacoustic emissions~\cite{Howard1988,Jacobs1990,Reichenbach2014,Martin2021,Roldn2021}. In mammals, two components most likely contribute to the active process on the cellular level: somatic motility and active hair-bundle oscillations. How these components precisely contribute to the four cardinal features of the ear is still debated. Here we use a complementary approach based on stochastic thermodynamics to identify distinct operational regimes of active hair-bundle oscillations that resemble signatures of different biological functions.
%. Two of these regimes suggest thermodynamic signatures of amplification and sensing, %respectively. Furthermore, our theory predicts that depending on the type of hair-bundle %oscillation, a cell can perform better as an amplifier or a sensor.

The nonlinear character of the hair-bundle's motion, as well as the nonequilibrium nature of its dynamics and fluctuations were previously analyzed using various stochastic models~\cite{Choe1998,Camalet2000,Martin2003,Vilfan2003,Nadrowski2004,Tinevez2007,Gelfand2010,Barral2010,Maoilidigh2012,Reichenbach2014,Bormuth2014,Bormuth2015,Barral2018,Belousov2020,Martin2021,Cao2021I,Cao2021II,Roldn2021,Tucci2022,Marcinik2024}. These studies were focused on principles, mechanisms, and properties of the observed dynamics. The hair bundle is thought to operate near a Hopf bifurcation undergoing limit-cycle oscillations, but also shows traits of bistability~\cite{Salvi2016}. Hair bundles' nonlinear mechanisms are associated with the cells' exquisite sensitivity and selectivity for transducing signals of particular frequencies and power. In bullfrogs---a paradigmatic model system in research on hair cells---the energy dissipated by the hair bundle per one oscillation cycle is estimated to be at least of the order of 100 $\kbt$ at the room temperature $T$, with Boltzmann constant $k_{\rm B}$. This amount of energy is comparable to the free energy released in hydrolysis of approximately $10$ molecules of adenosine triphosphate~(ATP) by myosin motors~\cite{Berger2022}, which presumably drive the motion of the hair bundle~\cite{Roldn2021}. However, it remains so far unknown how this consumed energy serves to implement the cells' functions, and how efficient are the concomitant thermodynamic processes.

The thermodynamic interpretation of earlier models encompasses interactions of the hair bundle with the environment as a heat reservoir, with an active feedback force controlled by a hidden state variable, and a second, active heat bath at an effective temperature, which represents spontaneous fluctuations associated with an internal adaptation mechanism~\cite{Tinevez2007,Nadrowski2004,Barral2018,Roldn2021}. Herein, we simplify the theoretical description by neglecting the second heat bath and retain only the deterministic features of the active dynamics. Notably, within such a minimalist approach we are able to fit experimental data with high accuracy, substantiating our quantitative analysis. On the other side, we incorporate into our model an external signal in the form of a time-periodic force, unveiling a rich landscape of distinct thermodynamic regimes characterized by the net direction of various energy flows within the system, as we discuss below. We refer rethermoaders to Refs.~\cite{Datta2022,Zhang2023,Davis2024,garcia2024optimal} for recent thermodynamics insights on periodically-driven and feedback-controlled minimal models active matter.

In this report we analyze the thermodynamic flows that result from interactions of a hair bundle with the thermal environment and two active agents---a feedback force and an external sinusoidal stimulus (Fig.~\ref{fig:intro}). Adopting the thermodynamic sign convention with positive net flow of heat or work from the environment or agent into the system, and negative otherwise, three components of the energy flows are considered.  First, the power $\dot{Q}$ absorbed from the environment as heat. Second, the active power $\dot{W}_{\rm a}$ supplied to the hair bundle by the hair cell. Third, the power $\dot{W}_{\rm e}$ extracted from the signal. Our analysis shows that, depending on the parameters of the system, the hair bundle can operate as an isothermal work-to-work machine with a feedback force in two distinct regimes. The power can be either extracted from or injected into the external signal by the hair cell. In the former case, which we call direct work transduction (DWT), the average active power is negative ($\avg{\dot{W}_{\rm a}} < 0$) and corresponds to an energy flow through the hair bundle into the cell, whereas the average external power is positive ($\avg{\dot{W}_e} > 0$, Fig.~\ref{fig:intro}). The second mode of operation is the reverse work transduction (RWT), characterized by the opposite sense of energy flow---from the cell to the external signal with positive active and negative external power. As we hypothesize, these two regimes might represent thermodynamic signatures of a hair cell's main functions: signaling and amplification.

Besides the work transduction between the two active agents, we also identify another nontrivial mode of operation---a \textit{feedback} refrigeration~\cite{Kim2007}, which consumes power of the external signal ($\avg{\dot{W}_{\rm e}} > 0$) in order to absorb heat ($\avg{\dot{Q}} > 0$) into the hair cell ($\avg{\dot{W}_{\rm a}} < 0$), thereby ``cooling'' the environment. This effect is thermodynamically viable due to the information flow exchanged between the active feedback force and the hair-bundle position~\cite{Sagawa2012,Munakata2014,Rosinberg2015,Rosinberg2016,Leighton2023,Leighton2024}. If not operating as a work transducer or refrigerator, the hair bundle falls back to the heater regime ($\avg{\dot{Q}} < 0$, $\avg{\dot{W_{\rm a}}} > 0$, $\avg{\dot{W_{\rm e}}} \ge 0$), as when there is no external signal present.

\begin{figure}[t!]%[tbhp]
\centering
\includegraphics[width=\columnwidth]{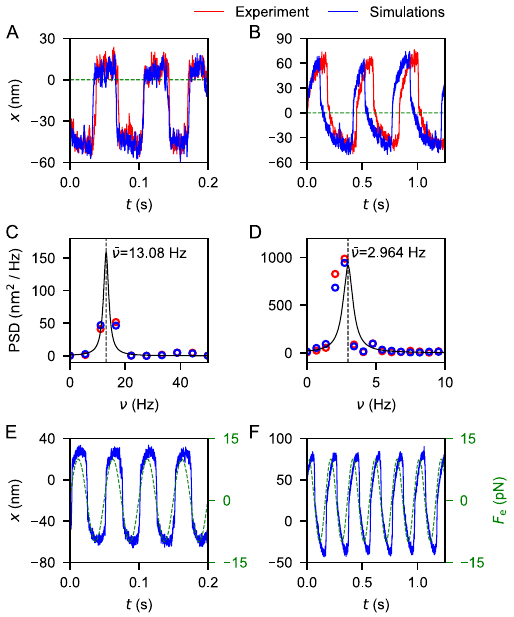}
\caption{Two examples of spontaneously oscillating hair bundles: Experimental time series of a hair-bundle's tip position $x$ (red) without an external stimulus in case \#1 (A), and case \#2 (B). Both are compared to simulations (blue) of Eqs.~[\ref{eq:1}]--[\ref{eq:motors}] with the parameter sets found by the simulation-based inference. Note that the phase of oscillations varies stochastically in each simulation of the model. Therefore, we report representative simulated trajectories with $t = 0$ chosen so as to match the initial phase of experiments. C, D: Power-spectrum density (PSD) calculated from the same data. To determine the natural frequency of oscillations $\bar{\nu}$ (dashed vertical line) we fit the Lorentzian profile (solid black curve) to the spectra obtained from simulated time series with a duration of $\SI{500}{s}$. E, F: Simulations of Eqs.~[\ref{eq:1}]--[\ref{eq:motors}] with an external sinusoidal force $F_{\rm e}$ (dashed green line) of amplitude $A_{\rm e} =\SI{10}{pN}$ and frequencies $\nu_{\rm e}=\SI{20}{Hz}$ (E, parameter set \#1) and $\nu_{\rm e} = \SI{5}{Hz}$ (F, parameter set \#2).}
\label{fig:oscillations}
\end{figure}

Using experimental data of hair-bundle oscillations from the bullfrog sacculus (Fig.~\ref{fig:oscillations}A and B, Materials and Methods), we identify a specific range of the external signal's frequencies and amplitudes, which make the hair bundle operate as a work-to-work machine. The DWT regime is viable when the signal amplitude exceeds a certain threshold and is most pronounced close to the natural frequency of the organelle's oscillations. These conditions suggest the presence of a threshold-gated sensing mechanism, which might trigger neuronal activity in the cell depending on the intake of the energy flow from the external stimulus. The RWT regime occurs at low signal amplitudes oscillating in a narrow range below the natural frequency of the hair bundle. Thus, a low-power stimulus could be amplified in a frequency-dependent manner by the internal adaptation process, which sustains oscillations of the organelle.

Our analysis of the experimental observations is based on a nonlinear parsimonious model \cite{Nadrowski2004,Tinevez2007,Barral2010,Maoilidigh2012,Belousov2020} and assisted by a simulation-based inference [SBI~\cite{tejero-cantero2020sbi,Cranmer2020,Tucci2022}]. By estimating model parameters of hair bundles oscillating in laboratory conditions at various frequencies, we predict their response to external signals of a sinusoidal form. We inspect in detail how the thermodynamic interactions of the system and its active environment depend on the frequency and amplitude of the external stimulation, and on other parameters of the model, which control the nonlinear regime of oscillations. Notably, we find that sharp self-sustained oscillations display better performance in the RWT mode, whereas smooth oscillations ensure robust DWT regime.

%%%
\section*{Nonlinear active dynamics of a hair bundle's oscillation}
In a laboratory, the spontaneous oscillations of a hair bundle's tip can be routinely observed by using a double-chamber preparation \cite{Martin1999,RamunnoJohnson2009,Azimzadeh2017,Azimzadeh2018,Alonso2020}. We recorded the time series of a hair bundle's position of cells from the bullfrog sacculus (Fig.~\ref{fig:oscillations}A and B, Materials and Methods), and selected four examples presenting regular oscillations at natural frequencies in the range \SIrange{1}{15}{Hz} for analysis.

To describe the time series of the hair bundle's tip position $x(t)$, we use a parsimonious model, which cathermon be derived from several models reported in the literature~\cite{Nadrowski2004,Tinevez2007,Barral2010,Maoilidigh2012} and which we refer to as a hidden Van der Pol -- Duffing oscillator~\cite{Holmes1980,SchenkHopp1996,Belousov2020,Belousov2019}, cf. \SIAppendix{}. The overdamped Langevin equation of motion for the hair bundle's position then reads
\begin{equation}\label{eq:1}
    \gamma \dot{x} = - \partial_x U(x,y)  + \Fe + \sqrt{2 \kbt \gamma} \zeta,    
\end{equation}
in which $U(x,y)$ is a potential generating a conservative force specified below. The tip of the hair bundle is subject to an external driving force $\Fe(t)$, which represents the signal, and to thermal noise of amplitude $\sqrt{2 \kbt \gamma}$ with friction coefficient $\gamma$; $\zeta$ is the standard Wiener noise of zero mean $\langle\zeta(t)\rangle=0$ and delta-correlated in time, $\langle\zeta(t)\zeta(s)\rangle = \delta(t-s)$. The derivative of the potential
\begin{equation}\label{eq:2}
    \partial_x U(x,y) = -A(x - y) + B(x - y)^3 + K_{\rm{sp}} x - C_x,
\end{equation}
includes a coupling to the internal active adaptation $y(t)$ and the following constants: coupling coefficients $A$ and $B$, a stiffness $K_{\rm{sp}}$, and a biFVanas term $C_x$. The dynamics of $y(t)$ is governed by
\begin{equation}\label{eq:motors}
    \dot{y} = -e y + b x + c_y,
\end{equation}
with constants $e$, $b$, and $c_y$. Equation~[\ref{eq:motors}] describes an internal active adaptation of the hair bundle to a new position. Such an adaptation has been shown to arise due to the interplay between mechanical forces, molecular motors, and calcium feedback, leading to experimentally-validated timescales on the order of a few milliseconds~\cite{Vilfan2003}.

By using the SBI approach~\cite{tejero-cantero2020sbi}, we determined values of the model parameters in Eqs.~[\ref{eq:1}]--[\ref{eq:motors}] from four experimental time series of hair-bundle oscillations at a natural frequency $\bar{\nu}$ in absence of the external driving ($\Fe\equiv0$, Fig.~\ref{fig:oscillations}A--D, SI~Fig.~\fgFitting). In the following, these four cases and the corresponding parameter sets are consistently numbered \#1--4. Whereas results for all four cases are reported in the \SIAppendix{}, here we focus on \#1 and \#2, which display sharp and smooth profiles of oscillations respectively. With the values of the model parameters inferred from experiments, we can probe the response of these hair bundles to a hypothetical external periodic signal
\begin{equation}\label{eq:signal}
    \Fe(t) = A_{\rm e} \sin(2\pi \nu_{\rm e} t)
\end{equation}
of amplitude $A_{\rm e}$ and frequency $\nu_{\rm e}$ by performing numerical stochastic simulations (Fig.~\ref{fig:oscillations}E and F).

\section*{Thermodynamics  of periodically-driven active hair-cell  bundles}
The physical model, as introduced above, allows us to apply systematically concepts of stochastic thermodynamics~\cite{seifert2012stochastic,peliti2021stochastic,roldan2022martingales}. In our setup, the hair bundle's tip position $x$ describes the state of the system that exchanges heat $Q$ with an isothermal bath at temperature $T$ (the extracellular medium). Furthermore, two external agents, the periodic external signal and the cell's active adaptation process, exert the work $\We$ and $\Wa$ on the system, respectively (Fig.~\ref{fig:intro}). Such a thermodynamic system is commonly known as an isothermal work-to-work machine with a feedback control~\cite{Sagawa2012,Munakata2014,Rosinberg2015,Rosinberg2016,Gupta2017,Leighton2023,Leighton2024}, as it is capable of transducing energy between the two external agents.

\begin{figure}[t!]
\includegraphics[width=\columnwidth]{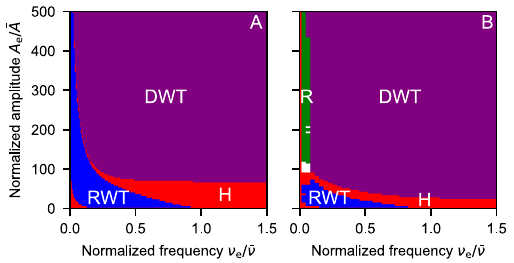}
\caption{Diagram of thermodynamic regimes for hair bundles described by the parameter sets \#1 (A) and \#2 (B): direct (DWT,  purple) and reverse (RWT, blue) work-to-work transduction, heater (H, red), and refrigerator (R, green). White-colored areas correspond to noisy regimes, in which fluctuations of energy flows are too large to assign one of the listed regimes with confidence. The amplitudes of the external signals are normalized by $\bar{A} = \sqrt{2 \kbt \gamma \bar{\nu}}$. The normalization constants take the value $\bar{A}_1 = \SI{0.12 \pm 0.01}{pN}$ and $\bar{A}_2 = \SI{0.111 \pm 0.001}{pN}$ with the natural frequencies $\bar{\nu}_1 = \SI{13.077 \pm 0.001}{Hz}$ and $\bar{\nu}_2 = \SI{2.9641 \pm 0.0004}{Hz}$ of the unperturbed oscillations described by the respective parameter sets.
}
\label{fig:thermo}
\end{figure}

Following the formalism of Sekimoto~\cite{sekimoto1998langevin}, the rate of heat absorption $\dot{Q}$ by the hair bundle from the heat bath in the time interval $[t, t + dt]$ is defined as
\begin{equation}\label{eqn:dQ}
    \dot{Q} \equiv (- \gamma \dot{x} + \sqrt{2 \kbt \gamma} \zeta) \circ \dot{x} = \left[\partial_x U(x,y) - \Fe \right] \circ \dot{x},
\end{equation}
in which we interpret $\circ$ as the Stratonovich-type product. The stochastic work $d\Wa$ exerted by the internal active adaptation process on the hair bundle is then
\begin{equation}\label{eqn:dWa}
    \dot{W}_{\rm a} \equiv \partial_y U(x,y) \circ \dot{y},
\end{equation}
whereas the work done by the external signal on the hair bundle reads
\begin{equation}\label{eqn:dWe}
    \dot{W}_{\rm e} \equiv  \Fe \circ \dot{x}.
\end{equation}
The above definitions, Eqs. [\ref{eqn:dQ}]--[\ref{eqn:dWe}], ensure that the sum 
\begin{equation}\label{eq:law1}
    \dot{Q} + \dot{W}_{\rm a} + \dot{W}_{\rm e} = \partial_x U(x,y) \circ \dot{x} + \partial_y U(x,y) \circ \dot{y}=\dot{U}
\end{equation}
yields the total change of internal energy and thereby fulfills the first law of stochastic thermodynamics at the single-trajectory level on the time interval $[t, t + dt]$~\cite{sekimoto1998langevin}.

The thermodynamic regime of the hair-bundle oscillatory state can thus be characterized by time averages of the heat flow $\avg{\dot{Q}}$ (dissipated energy), of the active power $\avg{\dot{W}_{\rm a}}$, and of the power supplied by the external signal $\avg{\dot{W}_{\rm{e}}}$, which all can be estimated numerically in simulations~(\SIAppendix{}). In the presence of a sinusoidal oscillatory signal $\Fe(t)$ of the form~[\eqref{eq:signal}], these net energy-flow rates become functions of the amplitude $A_{\rm e}$ and the frequency $\nu_{\rm e}$ of the external stimulus.

\begin{figure*}[t!]
\centering
\includegraphics[width=\textwidth]{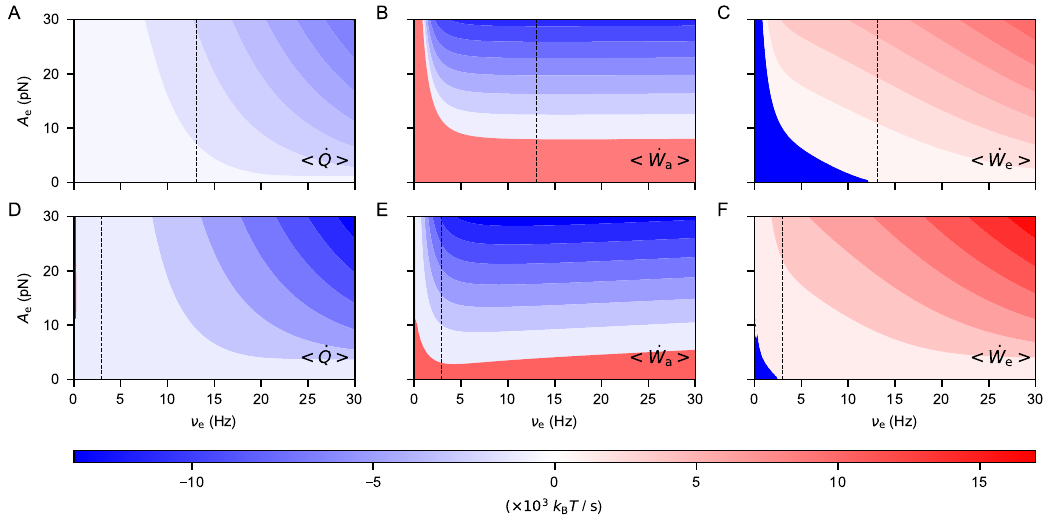}
\caption{Thermodynamic energy flows of an externally stimulated hair-bundle as functions of the stimulus' frequency $\nu_e$ and amplitude $A_{\rm e}$, characterized by the parameter sets \#1 (A--C) and \#2 (D--F): heat dissipation rate (A, D), active power (B, E), and external power (C, F). The vertical dashed line indicates the natural frequency of the unperturbed hair bundle $\bar{\nu}$. The respective energy flows are determined by integrating numerically Eqs.~[\ref{eqn:dQ}]--[\ref{eqn:dWe}] with a simulation time step of \SI{0.1}{ms} and averaging over \SI{500}{s}.
}
\label{fig:heatmaps}
\end{figure*}

From our numerical results, we systematically identified four distinct thermodynamic regimes in the space of parameters $A_{\rm e}$ and $\nu_e$  associated with different sign combinations of $\avg{\dot{Q}}$, $\avg{\dot{W}_{\rm a}}$, and $\avg{\dot{W}_{\rm e}}$ (Fig.~\ref{fig:thermo}, SI~Fig.~\fgMap). To categorize these regimes of our system for a specific set of parameters, we require that the three averaged energy flows, $\avg{\dot{Q}}$, $\avg{\dot{W}_{\rm a}}$, and $\avg{\dot{W}_{\rm e}}$ measured from $500$-\si{s} time series of simulations are greater than or equal to the standard error of their means. Close to the boundaries between the different regions, the energy flows fluctuate so much that a thermodynamic classification of such steady oscillatory states becomes impractical or would require very long simulations.

\subsubsection*{Heater regime}
For the experimental conditions of spontaneous oscillations ($F_{\rm e} \equiv 0$, $\avg{\dot{W}_{\rm e}} \equiv 0$), we find $\avg{\dot{W}_{\rm a}} > 0$ and $\avg{\dot{Q}} < 0$, which amount to active work done by the cell on the hair bundle and dissipated into the heat bath. When $F_{\rm e}\not\equiv 0$, the hair bundle is exposed to a harmonic signal, which typically carries a total energy proportional to $\nu_{\rm e}^2 A_{\rm e}^2$. The effect of this external driving on the energy flows in the system has a non-trivial dependency on the amplitude $A_{\rm e}$ and frequency $\nu_{\rm e}$~(Fig.~\ref{fig:heatmaps}, SI~Fig~\fgFlows).

For frequencies $\nu_{\rm e}$ of the signal which exceed the natural frequency of the hair bundle $\bar{\nu}$, the rate of heat exchanged with the environment becomes progressively more negative as the total power carried by the signal grows with $\nu_{\rm e}$ and $A_{\rm e}$ in all the four sets of parameter values~(Fig.~\ref{fig:heatmaps}A and D, SI~Fig.~\fgQ). This trend reflects the increase of the dissipated power $\avg{\dot{W}_{\rm a} + \dot{W}_{\rm e}}$ because of the additional external source of work $\avg{\dot{W}_{\rm e}} > 0$ up to a 100-fold increase of the entropy-production rate $\dot{S} = -\avg{\dot{Q}}/T$ in comparison with the unperturbed spontaneous oscillations of hair bundles~\cite{Roldn2021}. In this regime of negative heat flow $\avg{\dot{Q}} < 0$, and positive applied powers $\avg{\dot{W}_{\rm e}} > 0$, $\avg{\dot{W}_{\rm a}} > 0$ the system operates as a heater.

\subsubsection*{Refrigerator regime}
Besides the above heater regime, the system may also act as a weak refrigerator with $\avg{\dot{Q}} > 0$, $\avg{\dot{W}_{\rm a}} < 0$, $\avg{\dot{W}_{\rm e}} > 0$ at very low frequencies $\nu_{\rm e}$, but sufficiently large amplitudes $A_{\rm e}$ (Fig.~\ref{fig:thermo}B, Fig.~\ref{fig:heatmaps}D, SI~Fig.~\fgMap{D}).  We found this regime only for smooth profiles of the parameter sets \#2 and \#4. In the cases \#1 and \#3, we observed a weakly suppressed heat dissipation, which however remained on average always negative ($\avg{\dot{Q}} < 0$), in a similar region of $A_{\rm e}$ and $\nu_{\rm e}$.

The active term $\dot{W}_{\rm a}$, collecting information $I$ on the system through the feedback, enables the refrigerator regime, which agrees with the generalized second law of thermodynamics $\avg{\dot{S}} > -\avg{\dot{I}}$~\cite{Sagawa2010,Sagawa2012,Munakata2014,Rosinberg2015,Rosinberg2016,Rana2016,Leighton2023,Leighton2024}. Due to the nonlinear nature of equations, an analytical form of the rate of information transfer $\avg{\dot{I}}$ is not available for our system. However, we can prove the viability of the refrigerator regime, by using an equivalent formulation of our system with a second heat bath, whose temperature tends to zero (\SIAppendix{}).

\subsubsection*{Work-transduction regimes}
Two other regimes, which seem more relevant to the biological function of hair bundles, are characterized by the negative heat $\avg{\dot{Q}} < 0$ and transduction of energy between the external signal and the cell. We call them the \textit{direct} and \textit{reverse} work-to-work machines. In the DWT mode the hair cell harvests a fraction of the signal energy $\avg{\dot{W}_{\rm e}} > 0$ as active power $\avg{\dot{W}_{\rm a}}$, whereas in the RWT mode a fraction of the active power $\avg{\dot{W}_{\rm a}} < 0$ is channeled into the signal $\avg{\dot{W}_{\rm e}} < 0$ (Fig.~\ref{fig:heatmaps}B, C, E, and F, SI~Figs.~\fgWa{} and \fgWe{}).

The efficiency of the work-to-work machines may be quantified by the fraction of the total supplied power that is being transduced,
\begin{equation}
    \eta_{\rm in} = - \avg{\dot{W}_{\rm a}} / \avg{\dot{W}_{\rm e}},\qquad
    \eta_{\rm out} = - \avg{\dot{W}_{\rm e}} / \avg{\dot{W}_{\rm a}}
    \label{eq:effs}
\end{equation}
for the direct and reverse modes, respectively. Because the time-averaged internal energy $\avg{U}$ is constant, by virtue of the first law~Eq.~[\ref{eq:law1}]
$$\avg{\dot{U}} = \avg{\dot{Q}} + \avg{\dot{W}_{\rm a}} + \avg{\dot{W}_{\rm e}} = 0.$$
Therefore, as per the definition of the work-transduction regimes $\avg{\dot{Q}} < 0$, these are \textit{bona fide} efficiencies that always satisfy $0 < \eta_{{\rm in}} < 1$ in DWT ($\avg{\dot{W}_{\rm e}} > 0$, $\avg{\dot{W}_{\rm a}} <0$,  $\avg{\dot{Q}} < 0$), and $0 < \eta_{{\rm out}} < 1$ in RWT ($\avg{\dot{W}_{\rm e}} < 0$, $\avg{\dot{W}_{\rm a}} >0$, $\avg{\dot{Q}} < 0$). The upper bound of efficiency, with \SI{100}{\%} of the applied power being transduced, which may be achieved when $\avg{\dot{Q}}$ vanishes (i.e. at equilibrium), is not observed in practice.

\begin{figure}[t!]
\includegraphics[width=\columnwidth]{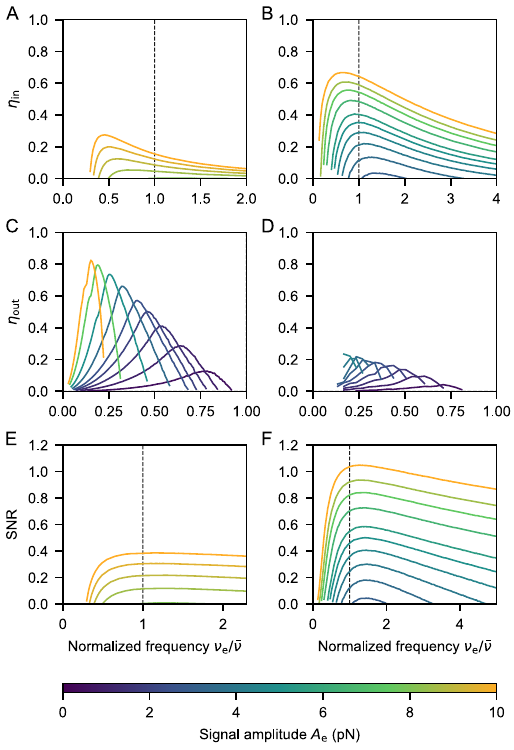}
\caption{Characterization of the hair bundles' work transduction by thermodynamic efficiencies (Eq.~[\ref{eq:effs}]) and by signal-to-noise ratio (SNR, Eq.~[\ref{eq:snr}]) for parameter sets~\#1 (A, C, E) and \#2 (B, D, F): the efficiency $\eta_{\rm in}$ of DWT (A, B), the efficiency $\eta_{\rm out}$ of RWT (C, D), and the SNR (E, F). While the hair bundle characterized by the parameter set~\#1 performs more efficiently in the RWT regime (C), the hair bundle characterized by the parameter set~\#2 has a larger efficiency of DWT (B) and displays a larger SNR (F). These differences suggest that cell~\#1 operates better as a sensor (DWT), while cell~\#2---as an amplifier (RWT).}
\label{fig:efficiency}
\end{figure}

Our numerical results for the efficiencies associated with the DWT and RWT modes of the parameter sets \#1 (Figs.~\ref{fig:efficiency}A, C) and \#2 (Figs.~\ref{fig:efficiency}B, D) reveal a non-monotonous behavior of both $\eta_{\rm in} $ and $\eta_{\rm out} $ as functions of the signal frequency. Within the explored parameter ranges, the peak efficiencies in general increase with the signal amplitude and may exceed \SI{80}{\%} of the supplied power. The location of their maxima depend on both the frequency and amplitude of the external signal (Fig.~\ref{fig:efficiency}A--D). At intermediate values of the amplitudes $A_{\rm e}$ the peak of $\eta_{\rm in}$ matches closely the natural frequency of the hair-bundle oscillation, shifting to the lower or higher frequencies for smaller or larger values of $A_{\rm e}$, respectively. The maximum of $\eta_{\rm out}$ occurs when the ratio $\nu_{\rm e}/\bar{\nu}$ is approximately one half, also moving towards lower or higher frequencies as the amplitude $A_{\rm e}$ increases or decreases. These observations confirm that efficiency and heat dissipation do not correlate in active systems, as suggested recently in Ref.~\cite{Baiesi2018}.

The parameter set \#1 is notably less efficient than \#2 in the DWT at small and moderate amplitudes $A_{\rm e} \le \SI{10}{pN}$ of the external signal (Fig.~\ref{fig:efficiency}A--B). Conversely, in the RWT regime the parameter set \#1 performs better than \#2 (Fig.~\ref{fig:efficiency}C--D). Parameters \#3 and \#4 compare similarly by their efficiencies $\eta_{\rm in}$ and $\eta_{\rm out}$ (SI Figs.~\fgIn--\fgSNR). As discussed in the next section, these differences can be attributed to distinct nonlinear regimes, in which the sets \#1, 3, and \#2, 4 operate.

The DWT requires that the signal amplitude exceeds a certain threshold depending on the model parameters, and is most pronounced when the stimulus frequency matches the natural frequency of the hair bundle. This tendency is further emphasized if we consider the extracted work $\avg{\dot{W}_{\rm a}}$ as the signal perceived by the cell and the standard deviation of the mean active power ${\rm Std}[\dot{W}_{\rm a}]$ as the uncertainty of the signal. We thus evaluate the signal-to-noise ratio as 
\begin{equation}
    {\rm SNR} = -\avg{\dot{W}_{\rm a}} / {\rm Std}[\dot{W}_{\rm a}],
    \label{eq:snr}
\end{equation}
which saturates close to the natural frequency of the hair bundle's oscillations (Fig.~\ref{fig:efficiency}E and F). 

The reverse regime of work transduction is limited to low frequencies $\nu_{\rm e} < \bar{\nu}$ and amplitudes $A_{\rm e}$ (Fig.~\ref{fig:efficiency}C and D). Assuming that the power carried by the signal is an increasing function of the signal and frequency, typically $\propto\nu_{\rm e}^2 A_{\rm e}^2$ for a harmonic oscillator, this regime is only triggered by and enhances weak stimuli.

In summary, our systematic analysis of the system's thermodynamics suggests four different operating regimes of a hair bundle, depending on the energetic flows induced by external signals of different amplitudes and frequencies. Our model fitted to experimental data revealed that distinct types of oscillations do not only influence the viability of these modes of operation, but also implicate different functions: one type of oscillation performs better as an amplifier---an RWT---and the other better as a sensor---a DWT. For a deeper understanding of the effects of the dynamical regimes, we next sought to characterize the thermodynamics close to their bifurcation points. We refer readers to very recent work highlighting the link between dissipation and the quality of stochastic oscillatory motion~\cite{Oberreiter2022,Santolin2025}.

\begin{figure*}[t!]
\centering
\includegraphics[width=\textwidth]{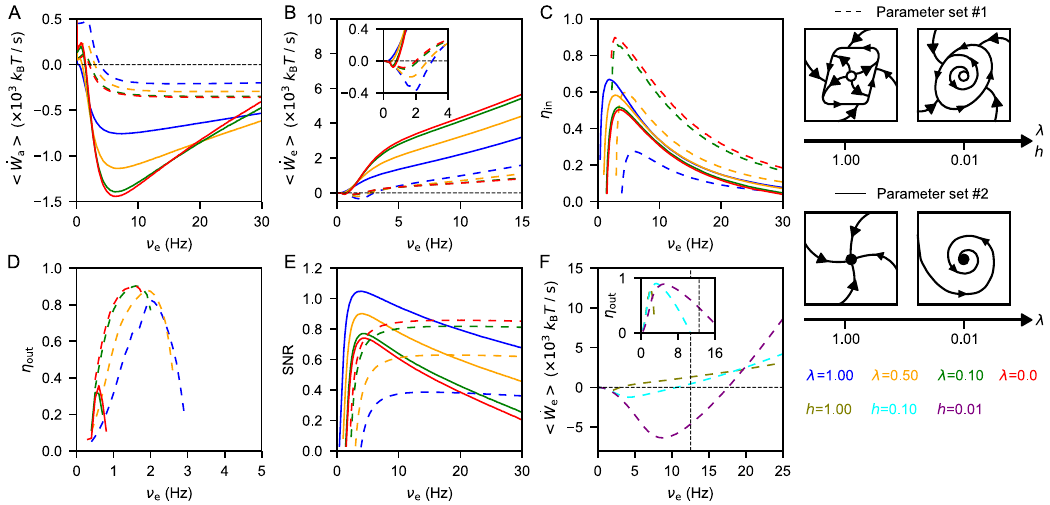}
\caption{Work transduction in different dynamical regimes: parameter sets \#1 (dashed curves, uninodal relaxation oscillations, approaching the Hopf bifurcation as $\lambda\to 0$ and the undamped harmonic oscillator as $h\to 1$) and \#2 (solid curves, monostabile regime, approaching the Hopf bifurcation as $\lambda\to0$) for the external signals of the same amplitude $A_{\rm e} = \SI{10}{pN}$ (A, B) and various frequencies $\nu_{\rm e}$.
A: While approaching the Hopf bifurcation ($\lambda \to 0$), curves describing the active power shift down in both cases, \#1 and \#2, implying more power intake in the direct work transduction and less power expanded in the reverse regime.
B: The curves of the external power become steeper in the proximity of the Hopf bifurcation of the parameter set \#2, indicating that more energy is consumed, but also more is transduced by the reverse work-to-work machine. The opposite trend is observed in the case \#1---the curves flatten.
C: Whereas the power intake of the monostable regime (\#2) increases closer to the Hopf bifurcation, the efficiency of work transduction decreases. The uninodal regime (\#1) transduces much less power, but more efficiently in the proximity of the Hopf bifurcation.
D: At a given amplitude of the external signal, the Hopf bifurcation enables reverse work transduction in the parameter set \#2, and improves its efficiency in the set \#1.
E: The signal-to-noise ratio is maximal in the monostable regime, far from the Hopf bifurcation as $\lambda \to 1$.
F: The low-friction and weak-nonlinearity limit of the Van der Pol -- Duffing Eq.~[\ref{eq:hvdf}] ($h\to0$, case \#1) resembles an underdamped harmonic oscillator. The reverse work transduction is maximal close to the half of the resonance frequency, which in the limit of $h\to0$ tends to $\nu_0 = \sqrt{c_1} / (2\pi)$ (dashed vertical line shows $\nu_0/2$). The sketches on the right illustrate how the phase portraits of the system change when the scaling parameters $\lambda$ or $h$ change, whose values are color-coded as shown in the legend.}
\label{fig:hopf}
\end{figure*}

%%%
\section*{Thermodynamic signatures of a Hopf bifurcation}
%%%
Hair bundles are believed to fulfill their functions by operating in proximity of a Hopf bifurcation~\cite{Ospeck2001,Hudspeth2010,Salvi2015}. Our experimental time series fall into two kinds of nonlinear regimes, out of the total four types available in our model (\SIAppendix{}). Namely, examples with sharp oscillatory profiles \#1 and \#3 belong to a \textit{uninodal} regime of relaxation oscillations, whose phase space contains a single unstable fixed point and a limit cycle. The models characterized by the parameter sets \#2 and \#4, which exhibit smooth oscillation profiles in their time series, fall into a particular kind of a \textit{monostable} regime---with a single stable fixed point,---because of the constant offset terms $C_x$ and $c_y$ in Eqs.~[\ref{eq:2}] and [\ref{eq:motors}]. In the absence of these terms, the system would be bistable and contain one saddle and two stable nodes. Instead, the latter nodes become imaginary for sufficiently large values of $C_x$ and $c_y$, but the remaining ``phantoms'' of these nodes distort the system's phase portrait. Although both, the monostable and bistable regimes, by themselves, do not display self-sustained relaxation oscillations~\cite{Holmes1980,SchenkHopp1996}, when driven by noise they may resemble a limit-cycle behavior, as the system slows down near the saddle nodes or their ``phantoms.''

Below we analyze how the thermodynamic flows change as the model's parameters approach a Hopf bifurcation. To this end, we consider an equivalent formulation of hair-bundle dynamics, which eliminates the variable $y$ by considering the difference $z = x - y$ (\SIAppendix{}). This formulation reveals a hidden Van der Pol -- Duffing oscillator $z(t)$ characterized by constants $c_0,c_1,c_2,c_3$, and $f_0$, and acting on the hair-bundle's position $x(t)$:
\begin{align}
    \label{eq:hvdf}
\ddot{z} &+ c_0\dot{z}+c_1 z+c_2 z^2\dot{z} + c_3 z^3 = f_0 \nonumber\\
&\quad + \frac{\dot{\zeta}(t)}{\gamma} + \frac{\dot{F}_{e}(t)}{\gamma} - (b-e)\left(\frac{\zeta(t)}{\gamma} + \frac{F_{e}(t)}{\gamma}\right), \\
    x &= \frac{1}{\gamma}\int_0^t ds\, g_x(t - s) \nonumber\\
        &\times\left[A z(s) - B z(s)^3 + C_x + F_{\rm e}(s) + \sqrt{2 k_{\rm B} T \gamma} \zeta(s)\right],
\end{align}
in which $g_x(t) = \exp(- K_{\rm sp} t / \gamma)$ is a Green function.

To approach the Hopf bifurcation, we fix the values of $c_1$, $c_2$, $c_3$, $f_0$, and scale the friction-like parameter $c_0 \to \lambda \hat{c}_0$ with respect to its original value $\hat{c}_0$ by the dimensionless factor $0 \leq \lambda \leq 1$, so that for $\lambda = 0$ the system is at the bifurcation point. To implement this scaling, we choose to adjust the values of the original model parameters $A$, $b$, $c$, $e$, and $c_y$, while keeping the rest fixed (SI Tables \tbScaleOne{} and \tbScaleTwo{}).

In the uninodal regime characterized by parameter set~\#1, approaching the Hopf bifurcation, as $\lambda \to 0$ , has little effect on the active and external power exchange (Fig.~\ref{fig:hopf}A--B). This minor effect however considerably improves $\eta_{\rm in}$ in both the DWT and RWT  modes (Fig.~\ref{fig:hopf}C--D): A minor increase of the power intake $|\avg{\dot{W}_{\rm a}}|$ in the direct regime is accompanied by a subtle decrease of the energy $|\avg{\dot{W}_{\rm e}}|$ extracted from the signal and leading to a drastic increase in the efficiency $\eta_{\rm in}$. In the RWT regime, the increase of $\eta_{\rm out}$ is less striking, as both power intake $|\avg{\dot{W}_{\rm a}}|$ and output $|\avg{\dot{W}_{\rm e}}|$ are decreasing, but the latter decreases slower with $\lambda\to0$.

Remarkably, with the set of parameters \#2, scaling by $\lambda \to 0$ enhances interactions with the active adaptation and external signal: both the active and external energy flows $\avg{\dot{W}_{\rm a}}$ and $\avg{W_{\rm e}}$ change severalfold (Fig.~\ref{fig:hopf}A--B). In the DWT regime, the increase of power intake $|\avg{\dot{W}_{\rm a}}|$ is afforded thanks to even more intense extraction of energy $|\avg{\dot{W}_{\rm e}}|$ from the external signal, resulting in a lower efficiency $\eta_{\rm in}$ (Fig.~\ref{fig:hopf}C). The proximity of the Hopf bifurcation also extends the range of reverse work transduction to higher levels of signal amplitudes than in the original set of parameters \#2 (Fig.~\ref{fig:hopf}D).

Comparing the two parameter sets \#1 and \#2, we notice that the SNR exceeds \SI{100}{\%} only in the second case, sufficiently far ($\lambda \to 1$) from the Hopf bifurcation (Fig.~\ref{fig:hopf}E). Overall the scaling by $\lambda$ seems to affect the direct work transduction much stronger than the reverse.
The reverse work transduction, which is triggered by oscillations of the external signal closely matching the half of the hair-bundle's natural frequency, resembles an antiresonance phenomenon. To highlight this similarity, we apply another scaling procedure by a dimensionless factor $h$: so that $c_0 = h \hat{c}_0$, $c_2 = h \hat{c}_2$, and $c_3 = h \hat{c}_3$ with respect to the original values $\hat{c}_{i=0,2,3}$ of the parameter set \#1. This scaling turns~\eqref{eq:hvdf} into an undamped harmonic oscillator as $h\to0$. To implement such a transformation we again adjust parameters $A$, $b$, $c$, $e$, and $c_y$, keeping all others fixed (SI~Table \tbScaleOne).

While approaching the regime of a weakly damped harmonic oscillator with $h \to 0$ from the uninodal regime of the parameter set \#1, the reverse work transduction of the hair bundle is enhanced by an order of magnitude (Fig.~\ref{fig:hopf}F). In this limit, the system's linear response converges to that of an undamped harmonic oscillator with a fundamental frequency $\nu_0 = \sqrt{c_1}/(2\pi)$ and amplitude $4\sqrt{c_2/c_0}$, which can be estimated by perturbative methods (\SIAppendix{}). Extraction of the power by $\Fe(t)$ then corresponds to the phenomenon of antiresonance, when $\nu_{\rm e} = \bar{\nu}_0 / 2$. Given a fixed amplitude $A_{\rm e}$ of the stimulus, with $h \to 0$ we indeed observe an emergent peak in $\avg{\dot{W}_{\rm e}}$ near $\nu_{\rm e} \approx \nu_0 / 2$ (Fig.~\ref{fig:hopf}F, SI-Fig.~\fgLast{F}).

In summary, we investigated the thermodynamic behavior of an equivalent description---a Van der Pol -- Duffing oscillator---near a Hopf bifurcation (Fig.~\ref{fig:hopf}A--E, SI.~Fig.~\fgLast{A--E}). We systematically determined how the energy flowing between the hair bundle, the active feedback mechanism, and the external signal contributes to the efficiency of work transduction in proximity of the bifurcation point. Interestingly, we found that the two dynamical regimes related to different experimental time series, behave differently. While in one dynamical regime, the efficiency of DWT is enhanced close to the bifurcation, in the other one, it is compromised. However, for both dynamical regimes, the RWT efficiency is improved by operating close to the bifurcation.

\section*{Discussion}
As we have shown, interactions between an active feedback mechanism and external periodic driving create a rich palette of four possible thermodynamic regimes in a stochastic model model of hair-cell bundle oscillations. These regimes are triggered by specific frequencies and amplitudes of the external stimulus. They include a basic mode of operation as a heater and a nontrivial one as a refrigerator, alongside the direct (DWT) and reverse work transduction (RWT), which might be related to the biological function of amplifying and sensing by hair cells.

The flow of energy $\langle\dot{W}_{\rm e}\rangle$ from the external force into the system turns negative in a narrow region of signals with low frequencies and amplitudes. That is, the intracellular active process can selectively power up and thus amplify such weak signals by exploiting the regime of RWT. This result, which is also supported by previous experimental observations that a hair bundle exerts work on an externally attached fiber oscillating at low frequencies and small amplitudes~\cite{Martin1999}, relates to the phenomenon of antiresonance. In agreement with Ref.~\cite{Martin1999}, the RWT appears viable only when the external driving oscillates below the natural frequency of the hair bundle in both, uninodal and monostable, regimes of our experimental time series. We expect that our results and ideas may be exploited in the emerging field of energy harvesting from the efficient rectification of fluctuations of small (e.g. nanosize) systems~\cite{Mitcheson2010,Wang2015,Shi2019}. 

Whereas an unperturbed hair bundle continually dissipates active work done on the system by the internal adaptation force, externally stimulated it is also capable of harvesting energy from signals of frequencies $\nu_{\rm e} \gtrsim \bar{\nu}$ and amplitudes $A_{\rm e} \gg \sqrt{2 \kbt \gamma \bar{\nu}}$. This regime of direct work transduction is suggestive of a hypothetical mechanism, which could trigger the neural activity of a sensing hair cell on a steady intake of power. Thereby, the signal detection does not necessarily imply a large inflow of energy but might rely on a high quality of the signal relative to its fluctuations, e.g. as characterized by the signal-to-noise ratio.

In our experimental data, we identified two dynamical regimes of a hair-bundle's oscillatory behavior: uninodal relaxation oscillations with a remarkable potential of signal amplification, and a monostable regime capable of harvesting energy with a high signal-to-noise ratio. In principle, hair cells might rely on the affinity of the two regimes to the DWT and RWT to fulfill their amplification and sensing functions respectively.

Moreover, like in some other biological systems~\cite{Graf2024}, the Hopf bifurcation associated with the above two dynamical regimes may affect the hair-cell function. While approaching the bifurcation point, the uninodal regime displays more efficient amplification capabilities and the monostable regime loses its high signal-to-noise ratio of the direct work transduction. The parameters of the active oscillatory mechanism could also be controlled by additional active processes within hair cells or by the accessory structures present in the hearing and balance-sensing organs. These additional mechanisms are capable of fine-tuning the target levels of power transduction. In particular, hair cells must be able to sustain and assimilate relevant amounts of transduced energy to avoid mechanical damage, which could be caused by interactions with excessively powerful sources.

\section*{Materials and Methods}
\subsubsection*{Recording oscillations of a hair bundle's tip in a two-chamber preparation}
A dissected mechanosensitive epithelium of a bullfrog’s sacculus was mounted between two chambers, one filled with perilymph and the other with endolymph, to mimic the physiological condition of the inner ear. Under these conditions, the hair bundles of healthy cells display spontaneous, robust oscillations, as reported previously~\cite{Martin1999,RamunnoJohnson2009,Azimzadeh2017,Azimzadeh2018,Alonso2020}.
We observed oscillating hair bundles under an upright microscope with differential-interference-contrast optics. Because slowly oscillating bundles are easy to identify visually, we directly projected a high-contrast image of such a hair bundle onto a dual photodiode. After low-pass filtering at \SI{4}{kHz}, the calibrated signal of the photodiode reported the bundle's position in time.

The experiments were conducted in accordance with the policies of The Rockefeller University’s Institutional Animal Care and Use Committee (IACUC Protocol 16942).

\subsubsection*{Simulation-based inference of model parameter values}
To identify parameter values of the model, which reproduce the experimentally recorded time series of the hair bundles' tips, we applied a simulation-based inference method~\cite{tejero-cantero2020sbi}. For each experimental case \#1--4, we generated $5 \times 10^6$ trajectories of the corresponding duration with parameter values sampled from a uniform prior, in order to train neural posterior estimators.

As the posterior density estimator, we chose a backend based on the normalizing flows. The parameter values were inferred by maximizing a posteriori probability of the trained estimator given summary statistics of the time series. Further implementation details are discussed in supporting information.

\section*{Acknowledgements}
Y.T. acknowledges financial support from the Royal Thai Government.
L.S. acknowledges financial support from the ICTP.
\'E.R. acknowledges financial support from PNRR MUR project PE0000023-NQSTI.  R.B. acknowledges funding from the EMBL.
The authors are grateful to Pascal Martin for stimulating discussions on amplification of external stimuli by hair bundles.
% Bibliography
\printbibliography

\cleardoublepage
\appendix
\onecolumn
\setcounter{equation}{0}
\setcounter{figure}{0}
\setcounter{table}{0}
%%%
\vspace{3em}
\begin{center}\huge\textbf{Supplemental Information}\end{center}
\vspace{-1em}

\renewcommand{\theequation}{S\arabic{equation}}
\renewcommand{\thefigure}{S\arabic{figure}}
\renewcommand{\thetable}{S\arabic{table}}

%%%
\section*{Hidden Van der Pol -- Duffing model of hair-bundle oscillations}
Our mathematical description of experimental data adapts to a parsimonious form of Ref.~\cite{Maoilidigh2012} and previously introduced models~\cite{Nadrowski2004,Tinevez2007,Barral2010,Barral2018}, which we extend with an additional force term $\Ka \Xa$ linearly dependent on the position of the adaptation motors $\Xa$ with a constant $\Ka$,
\begin{align}\label{eq:position}
\gamma \dot{X} &= - \Kg(X- \Xa - D P_{\rm o})- \Ks X + \Fe(t) + \zeta(t),
\\\label{eq:motor} \ga \dot{X}_{\rm a} &= \Kg(X - \Xa - D P_{\rm o}) - \Fm (1-SP_{\rm o}) - \Ka \Xa + \zeta_{\rm a}(t).
\end{align}
Here $X$ and $\Xa$ are the position of the hair bundle and of the adaptation motors respectively, $\zeta(t)$ and $\zeta_{\rm a}(t)$ are independent, delta-correlated in time, Gaussian white-noise terms with zero mean, $\Fe$ is the external force from a signal, and $P_{\rm o}$ is the probability of transduction channels being open. The dots above the variables indicate the time derivatives, whereas the constant parameters of the original model are explained in \Tab{tab:parameters1}. Furthermore the probability
\begin{equation}\label{eq:oprop} 
P_{\rm o} = \left[
    1 + \exp\left(
        \frac{N \Delta G + \Kg D^{2} / 2 - \Kg D (X - \Xa)}{N \kBT}
    \right)
\right]^{-1},
\end{equation}
is derived from the Boltzmann distribution for $N$ transduction elements with $\Delta G$ being the intrinsic energy difference between open and closed channels' states~\cite{Barral2010}.

\begin{table}[htbp!]\centering
\caption{\label{tab:parameters1}Range of parameter values for Eqs.~[\ref{eq:position}] and [\ref{eq:motor}] from Ref.~\cite{Barral2010}.}
\begin{tabular}{cll}
    \toprule
    Parameter & Definition & Range of values \\
    \midrule
    $\gamma$ & Friction of hair bundle & \SIrange{0.6}{5.3}{\mu N\cdot s\cdot m^{-1}} \\
    $\ga$ & Friction of adaptation motors & \SIrange{1.4}{21.9}{\mu N\cdot s\cdot m^{-1}} \\
    $\Kg$ & Combined gating-spring stiffness & \SIrange{0.37}{0.83}{m N\cdot m^{-1}} \\
    $\Ks$ & Combined pivot stiffness & \SIrange{0.08}{0.70}{m N\cdot m^{-1}} \\
    $D$ & Gating-swing of a channel & \SIrange{51}{71}{n m} \\
    $\Fm$ & Maximal motor force & \SIrange{37}{70}{p N} \\
    $S$ & Calcium feedback strength & \SIrange{0.29}{1.06}{} \\
    $N$ & Number of transduction elements & \SI{50}{} \\
    $\Delta G$ & Intrinsic free energy difference & \SI{10}{\,}$\kBT$ \\
    \bottomrule
\end{tabular}
\end{table}

To facilitate the connection between modeling and experimental data, we simplify the above description. First, we neglect one source of noise and set $\zeta_{\rm a}(t)/\ga \to 0$ by assuming the large-friction $\ga$ limit. The term $\zeta(t)$ describes the thermal noise from the environment and thus its strength is given by the fluctuation-dissipation theorem
$\avg{ \zeta(t) \zeta(s) } = 2 \kBT \gamma \delta(t-s)$. For any experimental measurement of the position $\Xe(t)$, we have to choose an origin of the reference frame. It does not in general correspond to $X = 0$ in Eqs.~[\ref{eq:position}] and [\ref{eq:motor}]. Therefore, the laboratory frame is shifted by a constant $\Xo = x - X$, implying the observable $\Xe = X + \Xo$ and a shifted value $\Xae = \Xa + \Xo$ of a hidden variable with equations of motion
\begin{align}\label{eq:shiftX}
\gamma \dot{\Xe} &= - \Kg(\Xe - \Xae - D P_{\rm o}) - \Ks (\Xe - \Xo) + \Fe(t) + \zeta(t),
\\\label{eq:shiftXa} \ga \dot{Y} &= \Kg(x - \Xae - D P_{\rm o}) - \Fm (1-SP_{\rm o}) - \Ka (\Xae - \Xo) + \zeta_{\rm a}(t).
\end{align}
Note that $P_{\rm o}(X - \Xa) = P_{\rm o}(\Xe - \Xae)$ remains invariant with respect to the shift of the reference frame by a constant $\Xo$.

Next, we reduce the nonlinearity of the model by expanding $P_{\rm o}$ around the most sensitive operational point with half-open probability and introducing a new variable
\begin{equation}
    Z \equiv C_1 - C_2 (\Xe - \Xae),
\end{equation}
in which
\begin{equation}\label{eq:c12}
C_1 \equiv \frac{2 N \Delta G + \Kg D^2 }{ 2 N \kBT},\qquad
C_2 \equiv \frac{\Kg D}{N \kBT},
\end{equation}
which allow rewriting \Eq{eq:oprop} as
\begin{equation}
    P_{\rm o} = (1+e^{-Z})^{-1} \approx \frac{1}{2} - \frac{1}{4} Z + \frac{1}{48} Z^3
\end{equation}
accurate up to the fourth power of $Z$ due to the vanishing even-order terms.
Using this approximation in Eqs.~[\ref{eq:position}] and [\ref{eq:motor}] together with the substitution
\begin{equation}
    y \equiv \Xae + \frac{C_1}{C_2},
\end{equation}
implying $\Xae = y - (C_1/C_2)$ and $Z = - C_2(\Xe - y)$, we obtain
\begin{align}\label{eq:p} 
\gamma \dot{x} =& - \Ks \Xe + \Kg \left[ \left(\frac{D C_2}{4} - 1 \right) (\Xe - y) - \frac{D C_2^3}{48} (\Xe - y)^3 + \frac{D}{2} - \frac{C_1}{C_2} \right ] + \Ks \Xo + \Fe(t) + \zeta(t),
\\
    \ga \dot{y} =& - \Ka y +  \left[\Kg + \frac{C_2}{4}(\Fm S - \Kg D)\right] (\Xe - y)
    + \frac{C_2^3}{48}(\Kg D - \Fm S)(\Xe - y)^3 \nonumber\\ 
    &+ \frac{1}{2}(\Fm S - \Kg D) + \frac{C_1}{C_2}(\Kg + \Ka) - \Fm + \zeta_{\rm a}(t).
\end{align}
To reduce the dynamical equation for the internal variable $y$ to the parsimonious form of Ref.~\cite{Maoilidigh2012} without terms of cubic order in $Z$, we assume that the term $\Fm S$ due to the maximal force generated by the molecular motors and modulated by calcium is balanced with the tension in the gating spring attributed to the opening of the ion channels $\Kg D$, so that $\Fm S = \Kg D$. This assumption leads to a linear equation valid up to $Z^4$:
\begin{equation}\label{eq:m}
    \ga \dot{y} = - \Ka y + \Kg (\Xe - y) + \frac{C_1}{C_2} (\Kg + \Ka ) - \Fm + \zeta_{\rm a}(t).
\end{equation}
Finally in the large-friction limit $\gamma_a \gg 0$ we can cast Eqs.~[\ref{eq:p}] and [\ref{eq:m}] into the following form
\begin{empheq}[box=\fbox]{align}\label{eq:xmain}
\dot{\Xe} &=-k \Xe + a( \Xe - y)-c( \Xe - y)^3 + c_x + \frac{\Fe(t)}{\gamma}+\frac{\zeta(t)}{\gamma}, \\ \label{eq:ymain} 
\dot{y}&=b \Xe - e y + c_y.
\end{empheq}
These two equations define our main mathematical model used in the main text. Together with \Eq{eq:c12} we identify the parameters
\begin{align}\label{eq:krel}
k &= \frac{\Ks}{\gamma}, \\
a &= \frac{\Kg}{\gamma}\left(\frac{\Kg D^2}{4 N \kBT} - 1\right), \\
c &= \frac{\Kg^4 D^4}{48 \gamma (N \kBT)^3 }, \\
c_x &= - \frac{N \Delta G}{ \gamma D} + \frac{\Ks \Xo}{\gamma},\\
\label{eq:brel}
b &= \frac{ \Kg}{\ga}, \\
\label{eq:erel}
e &= \frac{\Kg + \Ka}{\ga}, \\\label{eq:cyrel}
c_y &= \left (\frac{N \Delta G}{ \ga \Kg D} + \frac{D}{2 \ga} \right )(\Kg + \Ka) - \frac{\Fm}{\ga}.
\end{align}
The relations~[\ref{eq:krel}]--[\ref{eq:cyrel}] connect parameters of our model to those described in \Tab{tab:parameters1} and help estimate a plausible range of their numerical values. Furthermore, the simplified Eqs.~[\ref{eq:xmain}] and [\ref{eq:ymain}] can be mapped to Eqs.~[1]--[3] of the main text with
\begin{equation}
    \Ks = \gamma k,\quad
    A = \gamma a,\quad
    B = \gamma c,\quad
    C_x = \gamma c_x.
\end{equation}

A property of Eqs.~[\ref{eq:xmain}] and [\ref{eq:ymain}], which is important for the fitting procedure, is that one may shift the constant $c_x$ by an arbitrary value, while keeping invariant the time derivatives and the terms depending on the difference $x - y$, including the nonlinear cubic term. Namely, a substitution $(x, y) = (x' + c', y' + c')$ shifts $c_x' = c_x - k c' $ and $c_y' = c_y + (b - e) c'$ in \Eq{eq:ymain}. One may even eliminate the constant offset entirely from Eq.~[\ref{eq:xmain}] by setting $c' = c_x / k$. However, as described above, we cannot determine from our experimental data the origin of such a coordinate system due to an arbitrary shift $\Xo$ of the laboratory reference frame. Therefore we must leave the constant $c_x$ as a nuisance parameter.

In general $c_y$ cannot be eliminated from the hidden-variable \Eq{eq:ymain}, while keeping the nonlinear term invariant. For example a substitution
$$(x,y) = \left(x'' - \frac{c_y}{b - e}, y'' - \frac{c_y}{b - e}\right),$$
becomes singular if $e \to b$, which would preclude the fitting procedure to vary parameter values in this admissible region.

At last we remark that Eqs.~[\ref{eq:xmain}] and [\ref{eq:ymain}] derived above generalize the hidden Van der Pol equation~\cite{Belousov2020}, as shown in the next section.

\section*{Nonlinear regimes of oscillations}
In this section we analyze nature of nonlinear regimes afforded by Eqs.~[\ref{eq:xmain}] and~[\ref{eq:ymain}], which provide other constraints on the parameter values of interest. By substituting $z = x - y$ and $y = x - z$ into these equations, we obtain respectively
\begin{align}\label{eq:s14} 
\dot{x} =&-kx+a z-c z^3+c_x+\frac{\Fe(t)}{\gamma}+\frac{\zeta(t)}{\gamma},
\\\label{eq:s15} 
\dot{x}=&\dot{z}+(b-e)x+e z+c_y.
\end{align}
By subtracting \Eq{eq:s14} from [\ref{eq:s15}] and taking its time derivative we relate $x$ and $\dot{x}$ to
\begin{align}\label{eq:posZ} 
(e-b-k)x =& \dot{z}+(e-a) z+c z^3-c_x+c_y-\frac{\zeta(t)}{\gamma}-\frac{\Fe(t)}{\gamma},
\\\label{eq:s17} (e-b-k)\dot{x} =& \ddot{z}+(e-a)\dot{z}+3c z^2\dot{z}-\frac{\dot{\zeta}(t)}{\gamma}-\frac{\dot{F}_{\rm e}(t)}{\gamma}.
\end{align}
Using Eqs.~[\ref{eq:posZ}] and~[\ref{eq:s17}] we eliminate $x$ and $\dot{x}$ from~[\ref{eq:s15}] to obtain
\begin{equation}\label{eq:s18} 
\ddot{z} + (k-a+e)\dot{z} + (ke+ab-ae) z + 3c z^2\dot{z} + c(e-b) z^3 = \frac{\dot{\zeta}(t)}{\gamma} + \frac{\dot{F}_{\rm e}(t)}{\gamma} - kc_y - (b-e)\left( c_x + \frac{\zeta(t)}{\gamma} + \frac{\Fe(t)}{\gamma}\right),
\end{equation}
which can be rewritten as Eq.~\eqHvdf
% \begin{equation}\label{eq:hvdf}
% \ddot{z}+c_0\dot{z}+c_1 z+c_2 z^2\dot{z} + c_3 z^3 = f
% \end{equation}
with constant coefficients
\begin{equation}\label{eq:coefs}
c_0 = k-a+e, \quad
c_1 = ke+ab-ae, \quad
c_2 = 3c, \quad
c_3 = c(e-b),
\end{equation}
and a time-dependent nonhomogeneous part $f(t) = f_0 + f_t$ represented by a constant offset term
$$
    f_0 = - k c_y - (b - e) c_x,
$$
and the time-dependent part
$$
    f_t = \frac{\dot{\zeta}(t)}{\gamma} + \frac{\dot{F}_{\rm e}(t)}{\gamma} - (b-e)\left(\frac{\zeta(t)}{\gamma} + \frac{\Fe(t)}{\gamma}\right).
$$

The homogeneous part, which is the left-hand side of Eq.~[\ref{eq:hvdf}], describes a hybrid Van der Pol -- Duffing oscillator $z(t)$. It gives the name to the model derived in the previous section as $x$ is coupled to $z$ in a system $(x, z)$ with equations of motion~[\ref{eq:s14}] and [\ref{eq:posZ}], or explicitly
\begin{align*}
    \dot{x} =&-kx+a z-c z^3+c_x+\frac{\Fe(t)}{\gamma}+\frac{\zeta(t)}{\gamma},\\
    \dot{z} =& (e-b-k) x - (e-a) z - c z^3 + c_x - c_y + \frac{\zeta(t)}{\gamma} + \frac{\Fe(t)}{\gamma}.
\end{align*}
This system reduces to the hidden Van der Pol oscillator~\cite{Belousov2020} as long as $c_3 = 0$ ($e = b$). This generalization is only possible thanks to the new term $K_a X_a$, which we introduced into \Eq{eq:motor}, because $e \to b$ as $K_a \to 0$ (Eqs.~\ref{eq:brel} and~\ref{eq:erel}).

The homogeneous Van der Pol -- Duffing oscillator $z(t)$ has four distinct dynamical regimes~\cite{Holmes1980,SchenkHopp1996,Belousov2019,Belousov2020}:
\begin{itemize}
  \item A monostable regime for $c_0 \geq 0$, $c_1 \geq 0$,
  \item A uninodal oscillatory regime for $c_0 < 0$, $c_1 \geq 0$ with one limit cycle encircling a unique unstable node in the phase space,
  \item A bistable regime $c_0 > c_1 c_2/c_3$, $c_1 < 0$,
  \item A multinodal oscillatory regime $c_0 \leq c_1 c_2/c_3$, $c_1 < 0$ with one limit cycle encircling three unstable nodes in the phase space.
\end{itemize}
If $c_2 \ne 0$ and $c_3\ne 0$, the system is globally stable only if $c_2 > 0$ and $c_3 > 0$. If $c_2 = 0$ and/or $c_3 = 0$, then the global stability requires $c_0 > 0$ and/or $c_1 > 0$ respectively. Note that the structure of regimes in the original model (Eqs.~[\ref{eq:position}] and [\ref{eq:motor}]) is more complex than for our simplified equations of motion.

The constant offset term $f_0$ may significantly change the phase portrait structure of the dynamics and complicates its analysis. In the report we investigate only the experimentally relevant monostable and uninodal regimes, as well as the Hopf bifurcation connecting the two. Therefore we summarize here the effect of the offset term in these regimes.

Both, monostable and uninodal regimes, are characterized by a single fixed point determined by the unique real root $Z_0(f_0, c_1, c_3)$ of a depressed cubic equation
\begin{equation}\label{eq:depressed}
    z^3 + \frac{c_1}{c_3} z - \frac{f_0}{c_3} = 0.
\end{equation}
This equation vanishes if its discriminant is negative:
$$
    \mathcal{D} = - \frac{4}{c_3^3} \left(c_1^3 + \frac{27}{4} f_0 c_3\right) < 0.
$$
Given that $c_2 > 0$ and $c_3 > 0$, the monostable and uninodal regimes are respectively distinguished by the negative and positive real part of the eigenvalues of the matrix
$$
    \begin{pmatrix}
    0 & 1 \\
    - c_1 - 3 c_3 Z_0^2 & - c_0 - c_2 Z_0^2
    \end{pmatrix}
$$
which implies uninodal self-sustained oscillations when $c_0 < - c_2 Z_0^2$, a Hopf bifurcation at $c_0 = -c_2 Z_0^2$, and a monostable regime otherwise.

%%%
\section*{Fitting experimental time series}
To identify parameter values which can reproduce our experimental time series, we use the simulation-based inference (SBI)~\cite{tejero-cantero2020sbi}. This machine-learning approach comprises three stages---model simulations, training, and inference---and requires a physical model of the system, which in our case consists of Eqs.~[\ref{eq:xmain}] and~[\ref{eq:ymain}], and a prior range of admissible parameter values.

A plausible range of parameters for Eqs.~[\ref{eq:position}]--[\ref{eq:oprop}], which were already published in Refs.~\cite{Barral2010,Barral2018}, are reported in Tables~\ref{tab:parameters1} and \ref{tab:parameters2}. Using Eqs.~[\ref{eq:krel}]--[\ref{eq:cyrel}] we adapt these values to our model. Because ranges of the parameters found in the literature may slightly differ, we combine the two sources~\cite{Barral2010,Barral2018} together (Table~\ref{tab:parameters3}). The range of parameter $\Xo$ is estimated from the minimum and maximum values of four experimental data cases used in this study. Moreover we expand admissible values of the parameter $\Delta G$ and assume a range of parameter $\Ka$ to be within the order of magnitude of $\Kg$. Through Eqs.~[\ref{eq:krel}] and~[\ref{eq:cyrel}] the values reported in Table~\ref{tab:parameters3} are translated into the corresponding range of eight fitting parameters of our model (Table~\ref{tab:parameters4}).

\begin{table}[t!]\centering
\caption{\label{tab:parameters2}Range of parameter values for Eqs.~[\ref{eq:position}] and [\ref{eq:motor}] from Ref.~\cite{Barral2018}, which splits $\gamma = \gamma_{\rm H}+ \gamma_{\rm C}$ into two distinct contributions.}
\begin{tabular}{cll}
    \toprule
    Parameter & Definition & Range of values \\
    \midrule
    $\gamma_{\rm H}$ & Hydrodynamic friction of the hair bundle & \SI{85}{\mu N\cdot s\cdot m^{-1}} \\
    $\gamma_{\rm C}$ & Additional friction of the hair bundle & \SIrange{0}{4915}{n N\cdot s\cdot m^{-1}} \\
    $\gamma_{a}$ & Friction of adaptation motors & \SIrange{2}{13}{mN\cdot s\cdot m^{-1}} \\
    $\Kg$ & Combined gating-spring stiffness & \SIrange{0.4}{0.8}{m N\cdot m^{-1}} \\
    $\Ks$ & Combined pivot stiffness  & \SIrange{0.1}{0.3}{mN\cdot m^{-1}} \\
    $D$ & Gating-swing of a channel & \SIrange{35}{63}{n m} \\
    $\Fm$ & Maximal motor force & \SIrange{43}{64}{p N} \\
    $S$ & Calcium feedback strength & \SIrange{0.4}{0.8}{}  \\
    $N$ & Number of transduction elements & \SIrange{40}{60}{}  \\
    $\Delta G$ & The intrinsic energy difference & \SI{10}{\,}$\kBT$ \\
    \bottomrule
\end{tabular}
\end{table}

\begin{table}[htbp!]\centering
\caption{\label{tab:parameters3}Range of parameter values for Eqs.~[\ref{eq:shiftX}] and [\ref{eq:shiftXa}] adapted from Refs.~\cite{Barral2010,Barral2018}.}
\begin{tabular}{cll} 
    \toprule
    Parameter & Definition & Range of values \\
    \midrule
    $\gamma$ & Friction of hair bundle & \SIrange{0.085}{5.3}{\mu N\cdot s\cdot m^{-1}} \\
    $\ga$ & Friction of adaptation motors & \SIrange{1.4}{21.9}{\mu N\cdot s\cdot m^{-1}} \\
    $\Kg$ & Combined gating-spring stiffness & \SIrange{0.37}{0.83}{m N\cdot m^{-1}} \\
    $\Ks$ & Combined pivot stiffness & \SIrange{0.08}{0.70}{m N\cdot m^{-1}} \\
    $D$ & Gating-swing of a channel & \SIrange{35}{71}{n m} \\
    $\Fm$ & Maximal motor force & \SIrange{37}{70}{p N} \\
    $S$ & Calcium feedback strength & \SIrange{0.29}{1.06}{} \\
    $N$ & Number of transduction elements & \SIrange{40}{60}{}  \\
    $\Delta G$ & Intrinsic free energy difference & \SIrange{5}{10}{}$\kBT$  \\
    $\kBT$ & Thermal energy at room temperature & \SI{4.1}{\pico N. \cdot \nano m}  \\
    $\Ka$ & Estimated coupling of position to adaptation motors & \SIrange{0}{10}{}$\Kg$  \\
    $\Xo$ & Estimated offset point of an oscillation & \SIrange{-40}{60}{n m} \\
    \bottomrule
\end{tabular}
\end{table}

\begin{table}[b!]\centering
\caption{\label{tab:parameters4} Range of parameter values for Eqs.~[\ref{eq:xmain}] and [\ref{eq:ymain}] calculated from \Tab{tab:parameters3} by using the relations~[\ref{eq:krel}]--[\ref{eq:cyrel}].}
\begin{tabular}{cl} 
    \toprule
    Parameter & Range of values \\
    \midrule
    $k$ & \SIrange{15.09}{8235.29}{s^{-1}} \\
    $a$ & \SIrange{-7759.65}{57927.3}{s^{-1}} \\
    $c$ & \SIrange{0.007}{670.12}{{nm}^{-2}\cdot s^{-1}} \\
    $c_x$ & \SIrange{-1156300}{358244}{nm \cdot s^{-1}} \\
    $b$ & \SIrange{16.90}{592.86}{s^{-1}} \\
    $e$ & \SIrange{16.90}{6521.43}{s^{-1}} \\
    $c_y$ & \SIrange{-32368.4}{639941.3}{nm \cdot s^{-1}} \\
    $\gamma$ & \SIrange{0.000085}{0.0053}{\pico N\cdot s\cdot {nm}^{-1}} \\
    \bottomrule
\end{tabular}
\end{table}

To reduce the computational cost of SBI, we narrow down the range of parameters' values to the oscillatory regimes identified by our dynamical analysis in the previous section. One combination of parameters, which consists of a triplet $k$, $a$, and $e$, can ensure that $c_0 < 0$ as found in all oscillatory regimes. Therefore the difference $k - a$ must be negative, given that $k > 0$ and $e > 0$ (\Eq{eq:coefs} and \Tab{tab:parameters4}). This conditions can be enforced by replacing $a = \alpha k$ with a dimensionless parameter $\alpha \ge 1$. In addition we leverage the fact that $e \ge b$ as per Eqs.~[\ref{eq:brel}] and [\ref{eq:erel}] (cf. \Tab{tab:parameters3}). Since $b$ and $K_a$ have positive values, $e$ also has a positive value and $e \geq b$. To this end we introduce another dimensionless parameter $\epsilon \ge 1$ instead of $e = \epsilon b$.

Because in the model simulations we use Ito-Euler method with a discrete time step $dt=\SI{10}{\mu s}$, the upper bound of parameters $k$, $a$, $b$, $c$, and $e$ has been truncated. Otherwise much smaller time steps would be required in the simulation stage, which make the SBI computationally very challenging. The lower bound of $k$, $b$, and $c$ was rounded to more convenient values as well.

The model parameters' range constrained as described above (Table~\ref{tab:parameters5}) was then used as a support of the multidimensional uniform prior distribution $p(\theta)$, from which we sampled five million sets of values $\theta = (k, \alpha, b, \epsilon, c, c_x, c_y, \gamma)$ for the simulation stage. For each case of experimental time series of duration $\tau_{\rm exp}$ we simulated a trajectory spanning a time interval $t \in [-\tau_{\rm exp}, \tau_{\rm exp}]$ and discarded the first half of the interval as a burn-in time for the system's relaxation to a steady profile of oscillations. From the second half of the trajectory we obtain a series $\left\{x(t_i)\right\}_{i=0}^K$ of length $K$ ($t_0 = 0$, $\tau_{\rm exp} = K \Delta{t} = t_K$) and equal steps $\Delta{t} = t_{i+1} - t_i$. Both, the experimental time series acquired with $dt = \SI{0.1}{ms}$, and the simulated ones with $dt = \SI{10}{\mu s}$, were effectively undersampled to a common resolution $\Delta{t} = \SI{1}{ms}$, which was sufficient for the SBI.

\begin{table}[htbp!]\centering
\caption{\label{tab:parameters5}Range of parameter values used in this study.}
\begin{tabular}{cl} 
    \toprule
    Parameter & Range of values \\
    \midrule
    $k$ & \SIrange{15}{1200}{s^{-1}} \\
    $\alpha$ & \SIrange{1}{15}{} \\
    $c$ & \SIrange{0.01}{30}{{nm}^{-2}\cdot s^{-1}} \\
    $b$ & \SIrange{1}{50}{s^{-1}} \\
    $\epsilon$ & \SIrange{1}{2}{} \\
    $c_y$ & \SIrange{-400}{400}{nm \cdot s^{-1}} \\
    $c_x$ & \SIrange{-40000}{40000}{nm \cdot s^{-1}} \\
    $\gamma$ & \SIrange{0.000085}{0.0053}{\pico N\cdot s\cdot {nm}^{-1}} \\
    \bottomrule
\end{tabular}
\end{table} 

The results of simulations for each individual set of parameter values $\theta$ are represented by a summary statistics $\chi[x(t)|\theta]$ extracted from the simulated time series. In the training stage of SBI~\cite{tejero-cantero2020sbi} a sequential neural posterior estimator, based on the normalizing flows, was used to learn the posterior distribution $p(\theta|\chi)$. The following quantities were extracted as the summary statistics:
\begin{enumerate}
  \item The power spectral density of $x(t)$ obtained by the Welch method with the Hamming window of size $L = 0.9 K$ and an overlap $0.99 L$ points. Only $\sqrt{L}$ lowest modes of the power spectrum were used in the summary statistics.
  \item The average $\bar{x}$ and the standard deviation $\sigma_x$ of the distribution of $x(t)$ sampled from the time series;
  \item The Hermite-function modes
    $$p_{mn} = \frac{1}{K}\sum_{i=0}^{K-1} h_m\left[\frac{x(t_i) - \bar{x}}{\sigma_x}\right] h_n\left[\frac{\Delta{x}(t_i)}{\sigma_x}\right]$$
    of the joint probability distribution $p(x, \Delta{x})$ with the forward finite differences $\Delta{x}(t_i) = x(t_{i+1}) - x(t_i)$ and Hermite functions $h_{m = 0,1,2,...,15}(x)$ of order $m$, e.g. as defined in Ref.~\cite{Tucci2022}.
\end{enumerate}

\begin{figure}[p!]
\centering
\includegraphics[width=1\columnwidth]{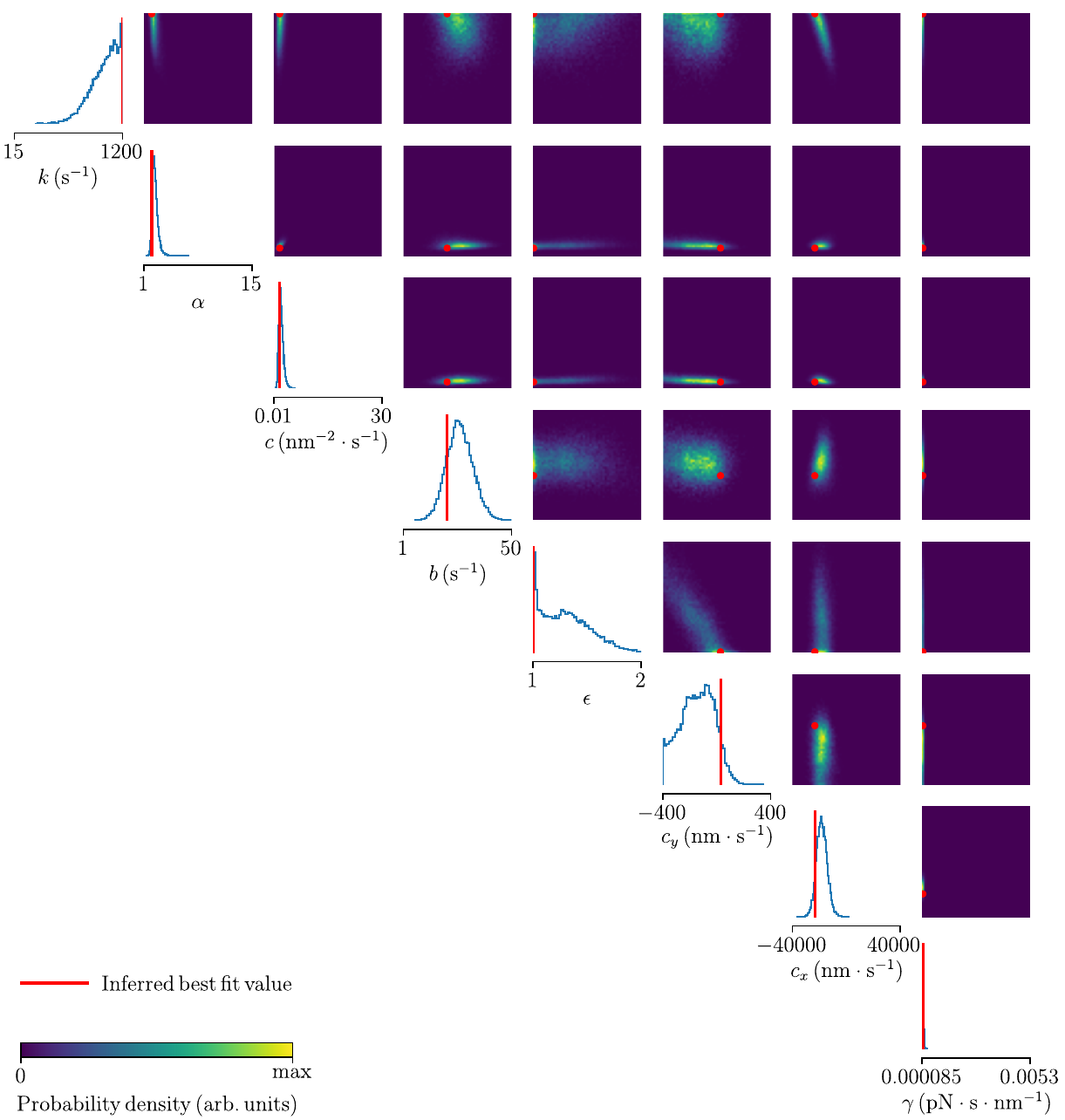}
\caption{Posterior distribution learnt from the simulation stage over the range of parameters in \Tab{tab:parameters5} with the time series of length $K = 201$ as in the experimental case \#1. The set of values inferred from these experimental time series by maximizing the posterior probability is shown in red (vertical lines in the 1D projections and red points in the 2D projections). The inferred best fit values are given in \Tab{tab:values}.} 
\label{fig:post_dist1}
\end{figure}

\begin{figure}[p!]
\centering
\includegraphics[width=1\columnwidth]{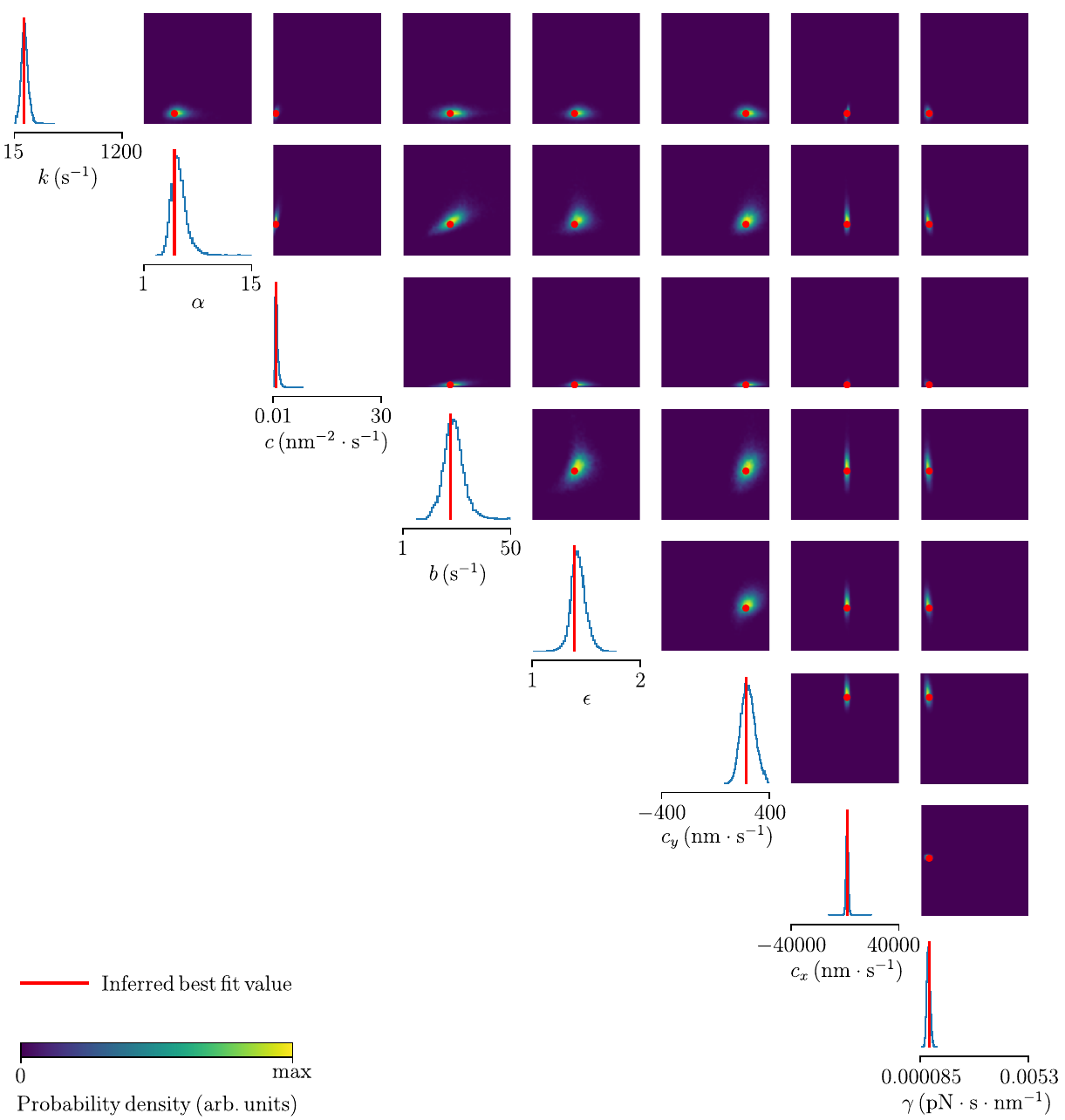}
\caption{Posterior distribution learnt from the simulation stage over the range of parameters in \Tab{tab:parameters5} with the time series of length $K = 1747$ as in the experimental case \#2. The set of values inferred from these experimental time series by maximizing the posterior probability is shown in red (vertical lines in the 1D projections and red circles in the 2D projections). The inferred best fit values are given in \Tab{tab:values}.}
\label{fig:post_dist2}
\end{figure}

\begin{figure}[p!]
\centering
\includegraphics[width=1\columnwidth]{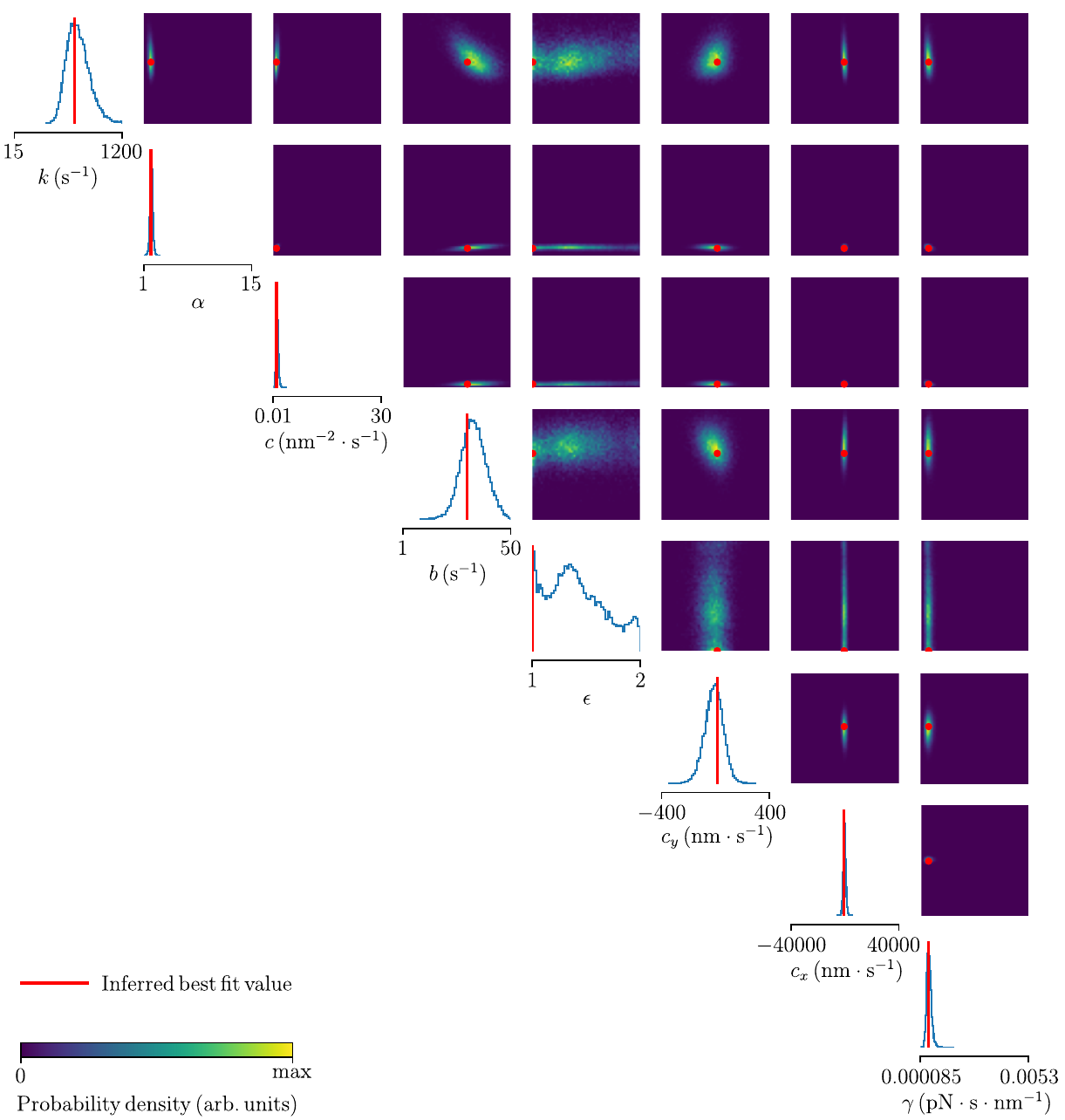}
\caption{Posterior distribution learnt from the simulation stage over the range of parameters in \Tab{tab:parameters5} with the time series of length $K = 212$ as in the experimental case \#3. The set of values inferred from these experimental time series by maximizing the posterior probability is shown in red (vertical lines in the 1D projections and red circles in the 2D projections). The inferred best fit values are given in \Tab{tab:values}.}
\label{fig:post_dist3}
\end{figure}

\begin{figure}[p!]
\centering
\includegraphics[width=1\columnwidth]{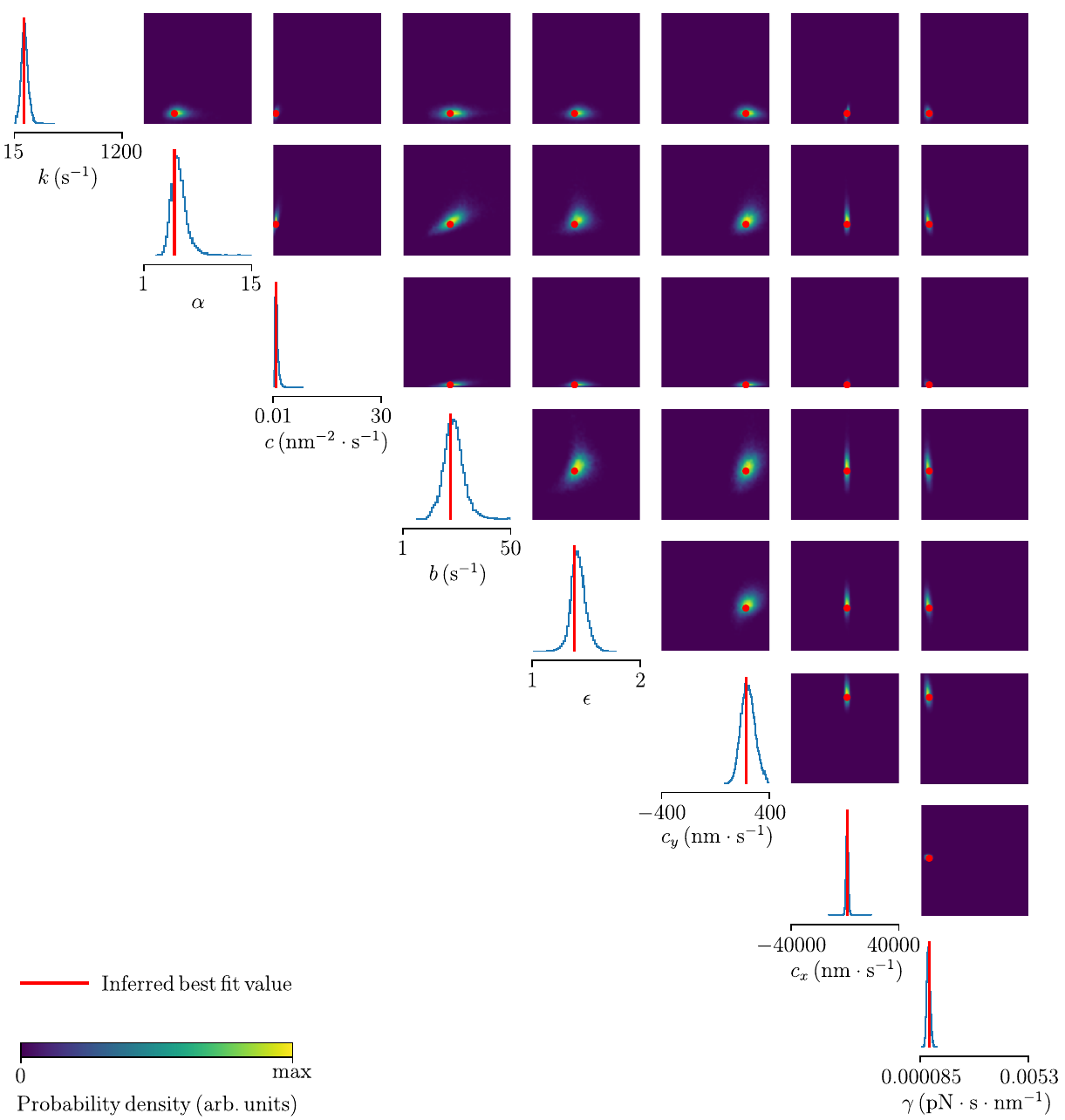}
\caption{Posterior distribution learnt from the simulation stage over the range of parameters in \Tab{tab:parameters5} with the time series of length $K = 1636$ as in the experimental case \#4. The set of values inferred from these experimental time series by maximizing the posterior probability is shown in red (vertical lines in the 1D projections and red circles in the 2D projections). The inferred best fit values are given in \Tab{tab:values}.}
\label{fig:post_dist4}
\end{figure}

Because the posterior distributions thus learnt were not all unimodal for the summary statistics extracted from the time series of the four experimental cases $\chi_{c = 1,2,3,4}$ (Figs.~\ref{fig:post_dist1}--\ref{fig:post_dist4}), in the inference stage we maximized the \textit{a posteriori} probability $p(\theta | \chi_c)$ with respect to the model parameters $\theta$. To estimate the uncertainties $\sigma(\hat{\theta}_i)$ of the optimal values $\hat{\theta}_{i=1,2,...,8}$ we evaluated the empirical Bayesian information matrix~\cite{Nguyen2017,Mentr1995} with elements given by
$$
    I_{ij} = \left\langle
        \frac{\partial \ln p(\theta | \chi_c)}{\partial\theta_i}\Big{|}_{\theta=\hat{\theta}}
        \frac{\partial \ln p(\theta | \chi_c)}{\partial\theta_j}\Big{|}_{\theta=\hat{\theta}}
    \right\rangle_{\theta},
$$
in which the average $\avg{\cdot}_{\theta}$ is taken over a sample of parameter values generated from the prior distribution $p(\theta)$. The standard errors of the fitted parameters then refer to the square roots of the diagonal entries in the inverse matrix $\sigma(\hat{\theta}_i) = \sqrt{(I^{-1})_{ii}}$.

\begin{figure}[p!]
\centering
\includegraphics[width=0.9\columnwidth]{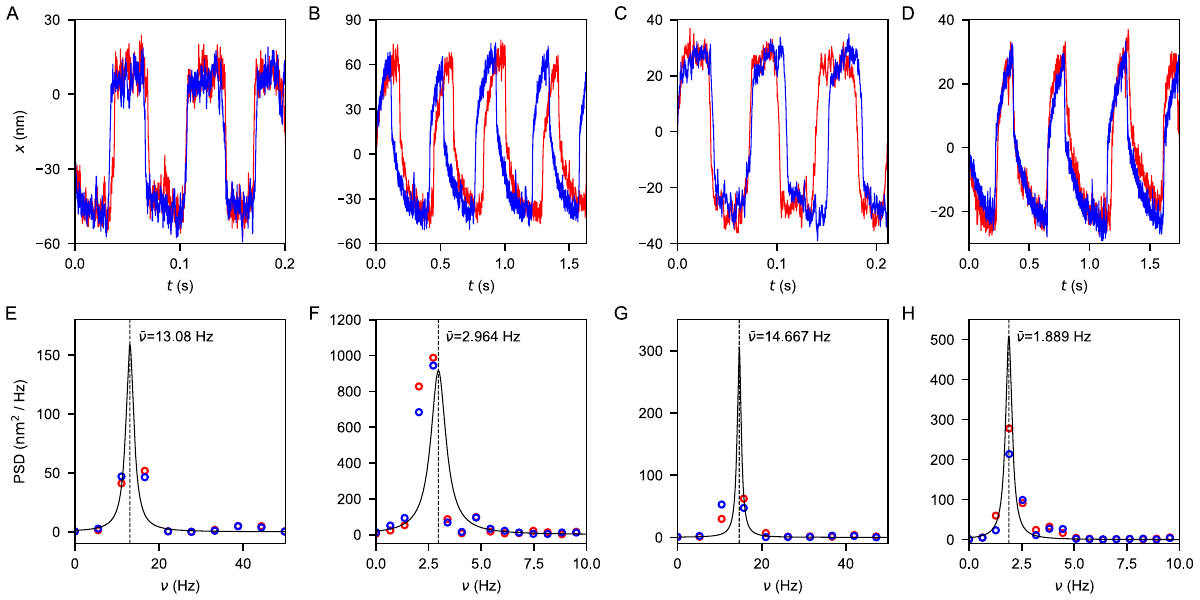}
\caption{Time series (A--D) and their power-spectra (E--H) from four experiments \#1--4 (red) are compared to the simulations (blue) performed with best-fit parameter values: \#1 (A, E), \#2 (B, F), \#3 (C, G), \#4  (D, H). Panel A, B, E, and F are also shown in Figure 2 of the maintext.}
\label{fig:time-series}
\end{figure}

The parameter values estimated by the SBI (\Tab{tab:values}) yielded an excellent agreement between the model simulations and experiments (Fig.~\ref{fig:time-series}). As described beforehand, the nuisance parameter $c_x$ is not well constrained and therefore has a large uncertainty (cf. Case 3 in \Tab{tab:values}). However $c_x$ and $c_y$ as additive constants of Eqs.~[\ref{eq:xmain}] and [\ref{eq:ymain}] should not contribute to the time-averaged thermodynamic energy flows $\dot{Q}$, $\dot{W}_{\rm a}$, and $\dot{W}_{\rm e}$.

\begin{table}[htbp!]\centering
\caption{\label{tab:values}Parameter values estimated for the four experimental time series by SBI. The uncertainties are specified by standard errors.}
\begin{tabular}{ccccc} 
    \toprule
    Parameter (unit) & Case 1 & Case 2 & Case 3 & Case 4 \\
    \midrule
    $k \, (\si{s^{-1}})$ & $1198.2\pm0.9$ & $124.4\pm2.0$ & $677.0\pm0.9$ & $136.9\pm1.7$ \\
    $a \, (\si{s^{-1}})$ & $2399\pm45$ & $619\pm12$ & $1300\pm12$ & $759\pm11$  \\
    $c \, (\si{{nm}^{-2}\cdot s^{-1}})$ & $1.68\pm0.08$ & $0.73\pm0.04$  & $0.914\pm0.025$ & $3.69\pm0.04$ \\
    $b \, (\si{s^{-1}})$ & $20.9\pm0.5$ & $22.59\pm0.12$ & $30.42\pm0.17$ & $15.12\pm0.07$ \\
    $e \, (\si{s^{-1}})$ & $21.0\pm0.5$ & $31.48\pm0.19$ & $30.60\pm0.24$ & $20.44\pm0.11$\\
    $c_y \, (\si{nm \cdot s^{-1}})$ & $28\pm8$ & $225.4\pm1.8$ & $13.1\pm2.9$ & $57.3\pm1.9$ \\
    $c_x \, (\si{nm \cdot s^{-1}})$ & $-23360\pm468$ & $1528\pm240$ & $-564\pm216$ & $264\pm550$ \\
    $\gamma \, (\si{\pico N\cdot s\cdot {nm}^{-1}})$ & $0.000127\pm0.000024$ & $0.000511\pm0.000011$ & $0.00048\pm0.000017$ & $0.001897\pm0.000016$ \\
    \bottomrule
\end{tabular}
\end{table}

%%%
\section*{Positive average heat flow}
The second law of thermodynamics for systems with a feedback mechanism~\cite{Sagawa2012,Munakata2014,Rosinberg2015,Rosinberg2016,Leighton2023,Leighton2024}, like ours, does not forbid the negative heat flow $\avg{\dot{Q}}$, because the active work $\avg{\dot{W}_{\rm a}}$ is capable of extracting entropy from the system. For linear feedback systems the corresponding information term can be derived in a closed form from a variant of the transient fluctuation theorem, but the cubic nonlinearity of the potential in our model makes a direct calculation unfeasible. However we may prove that in our model the entropy may flow from the heat bath into the system and then harvested in the form the active power, as observed in the refrigerator regime of our model.

To this end we redo our thermodynamic analysis, starting with Eq.~[\ref{eq:p}] and [\ref{eq:m}] cast as
    \begin{align}\label{eq:pos}
        \gamma \dot{x} &= - \partial_x V(x, y) + \Fe(t) + \zeta(t),
    \\\label{eq:mot}
        \ga \dot{y} &= - \partial_y V(x,y) - F_{\rm a}(x,y) - \Fm + \zeta_{\rm a}(t),
\end{align}
in which we do not neglect the noise $\zeta_{\rm a}(t)$ attributed to a \textit{second, active, heat bath} and consider the potential
$$
    V(x, y) = U(x, y) + \Ka y - C_y y = - \frac{A}{2}(x - y)^2 + \frac{B}{4}(x - y)^4 + \frac{\Ks}{2} x^2 - C_x x - C_y y + \frac{\Ka}{2} y^2
$$
as the system's internal energy with $C_y = \ga c_y$, whereas the term
$$
    F_{\rm a}(x, y) = \frac{\Kg D^2}{4 N \kBT} (x - y)
$$
represents the nonconservative active force generated by the motors. Then following the formalism of Ref.~\cite{Roldn2021} we define the heat exchanged by the system with the the active bath as
$$
    \dot{Q}_{\rm a} = (- \ga \dot{y} + \sqrt{2 \kBT \ga} \zeta_{\rm a}) \circ \dot{y}
        % = \left[\partial_y V(x, y) + F_{\rm a} \right]  \circ dy(t)
        = \partial_y V(x, y) \circ \dot{y} - \dot{W}_{\rm h},
$$
in which we recognize $\dot{W}_{\rm h} = - F_{\rm a} \circ \dot{y}$ as the hidden active work. Thereby the internal energy $V(x,y)$ observes the first law of thermodynamics:
$$
    \dot{Q} + \dot{W}_{\rm e} + \dot{Q}_{\rm a} + \dot{W}_{\rm h} = \partial_x U + \partial_y V = \partial_x V + \partial_y V = dV,
$$
with the heat $\dot{Q}$ and the work $\dot{W}_{\rm e}$ coinciding with our definitions in the main text.

With the above definitions the second law requires that the entropy change
\begin{equation}\label{eq:dS}
    dS = -\frac{dQ}{T} - \frac{dQ_{\rm a}}{T_{\rm a}} \ge 0,
\end{equation}
with the temperature $T_a$ of the active bath. In other words the second law allows the heat $dQ$ to flow into the system and then as $dQ_{\rm a}$ into the active bath.

As we take the limit $T_{\rm a} \to 0$ the two-baths model converges to Eqs.~[1]--[3] from the main text. Thereby we recognize that the average active work in our model
$$\avg{\dot{W}_{\rm a}} = \avg{\dot{Q}_{\rm a} + \dot{W}_{\rm h} + \dot{V} - \dot{U}} = \avg{\dot{Q}_{\rm a}} + \avg{\dot{W}_{\rm h}}$$
is the sum of two contribution introduced in this section. However, since $\avg{\dot{W}_{\rm a}}$ contains an implicit heat-flow term $\avg{\dot{Q}_{\rm a}}$ into the active bath with an arbitrarily low temperature, the second law \Eq{eq:dS} allows negative values of $\avg{\dot{W}_{\rm a}}$ simultaneously with a positive contribution of $\avg{\dot{Q}}$.

%%%
\section*{Linear-response theory for uninodal relaxtaion oscillations}
To analyze the system's response to an external perturbation in the regimes of uninodal relaxation oscillations, we first recognize that Eq.~[\ref{eq:s14}] is linear in $x$ and, thus, can be solved by
\begin{equation}\label{eq:green}
    x = \frac{1}{\gamma}\int_0^t ds\, g_x(t - s) \left[A z(s) - B z(s)^3 + C_x + F_{\rm e}(s) + \sqrt{2 \kBT \gamma} \zeta(s)\right],
\end{equation}
given that we have a solution of Eq.~[\ref{eq:hvdf}]. Below we seek an approximation of such a solution in the form
\begin{equation}\label{eq:volterra}
    z(t) \simeq z_0(t) + \int_0^t ds\, g_z(t - s) f_t(s),
\end{equation}
in which we introduce the first-order Volterra kernel $g_z(t)$---the linear response function of the inhomogeneous Van der Pol -- Duffing oscillator,---and neglect the higher-order contributions of the time-dependent perturbation $f_t(t)$ to the limit-cycle term $z_0(t)$, which we find first by setting $f_t(t) \equiv 0$~\cite{Belousov2019,Belousov2020}.

\subsection*{Lindstedt-Poincare approximation of the limit-cycle solution}
A dimensionless form of Eq.~[\ref{eq:hvdf}] without the time-dependent forcing $f_t(t) \equiv 0$ immensely simplifies the analysis of the limit-cycle solution $z_0(t)$. The textbook approach to the Van der Pol equation~\cite[Secs. 4.4 and 5.9]{jordan2007nonlinear} works well for the uninodal regime of the hybrid Van der Pol -- Duffing oscillator when $c_1 > 0$, as in the experimental cases \#1 and \#2 (an alternative parametrization is possible for $c_1 \le 0$, which is not relevant for our experiments): by using $\sqrt{-c_0 / c_2}$ as the unit of length, $(\omega \sqrt{c_1})^{-1}$ as the unit of time, we obtain
\begin{equation}\label{eq:uni}
    \omega^2 q''(t_\omega) + q(t_\omega) - \mu_1 \omega q'(t_\omega) \left[1 - q(t_\omega)^2\right] + \mu_2 q(t_\omega)^3 - \mu_3 = 0,
\end{equation}
where the prime denotes differentiation with respect to a new independent variable $t_\omega = \omega \sqrt{c}_1 t$, parameterized by the yet unknown frequency $\omega$, and a new dependent variable $q(t_\omega)$ is substituted for $z(t) = \sqrt{-c_0/c_2} q(t_\omega)$. Note that $q$, $\omega$, and $t_\omega$ are dimensionless. We also introduced two parameters $\mu_1 = -c_0/\sqrt{c_1}$ and $\mu_2 = -c_0 c_3 / (c_1 c_2)$, which control the nonlinearity strength associated with the Van der Pol and Duffing families of equations respectively, and a third parameter $\mu_3 = \sqrt{-c_2 / c_0} f_0 / c_1$ which represents a redimensionalized constant offset.

In the Lindstedt-Poincare method we approximate the limit-cycle solution of Eq.~[\ref{eq:uni}] subject to the initial-value conditions $q(0) = \alpha$ and $q'(0) = 0$ by positing multivariate power-series expansion of
\begin{align}\label{eq:lpmethod}
    &q(t_\omega) = q_0(t_\omega) + \mu_1 q_{100}(t_\omega) + \mu_2 q_{010}(t_\omega) + \mu_3 q_{001}(t_\omega) + O[(\mu_1 + \mu_2 + \mu_3)^2],\\
    &\omega = 1 + \mu_1 \omega_{100} + \mu_2 \omega_{010} + \mu_3 \omega_{001} + O[(\mu_1 + \mu_2 + \mu_3)^2],
\end{align}
which we truncate after the first-order terms.

By plugging Eq.~[\ref{eq:lpmethod}] into [\ref{eq:uni}] and collecting terms of equal powers in $\mu_{1,2,3}$, we obtain the following equation for $q_0(t_\omega)$:
$$
    q_0'' + q_0 = 0,
$$
which is solved by $q_0 = \alpha \cos t_\omega$. This intermediate result is however insufficient to constrain $\alpha$ and requires analysis of the equations for $q_{100}$, $q_{010}$, and $q_{001}$. The first-order terms of Eq.~[\ref{eq:uni}] in $\mu_{1,2,3}$ yield
\begin{align}\label{eq:first1}
    &q_{100}'' + q_{100} = q_0' - q_0' q_0^2 - 2 \omega_{100} q_0'',
    \\\label{eq:first2}
    &\ddot{q}_{010} + q_{010} = -q_0^3 - 2 \omega_{010} q_0'',
    \\\label{eq:first3}
    &\ddot{q}_{001} + q_{001} = 1 - 2 \omega_{001} q_0'',
\end{align}
which are all solved by a sum of the complementary solution and the convolution of the right-hand sides, further denoted $f_{100}(t_\omega)$, $f_{010}(t_\omega)$, and $f_{001}(t_\omega)$, with the green function $g(t_\omega) = \sin t_\omega$:
\begin{equation}
    q_{\text{idx}} = \alpha_{\text{idx}}\cos t_\omega + \beta_{\text{idx}}\sin t_\omega + \int_0^{t_\omega} ds\, g(t_\omega - s) f_{\text{idx}}(s)
\end{equation}
where $\alpha_{\text{idx}}$ and $\beta_{\text{idx}}$ are constants, and the index $\text{idx}$ stands for one of the values $100$, $010$, $001$.

Using the solution for $q_0(t_\omega)$ in $f_{\text{idx}}(t_\omega)$ we obtain the general approximation Eq.~[\ref{eq:lpmethod}], in which all coefficients $\alpha_{\text{idx}}$ and $\beta_{\text{idx}}$ must vanish to satisfy the imposed boundary condition $q_0(0) = \alpha$ and $q_0'(0) = 0$:
\begin{multline}
    q(t_\omega) \approx \mu_3 + \left[
        \alpha
        + \frac{\alpha \mu_1 t_\omega}{8} \left(4 - \alpha^2\right)
        - \frac{\alpha^3\mu_2}{32} - \mu_3
    \right] \cos t_\omega
    + \frac{\alpha^3 \mu_2}{32}\cos(3 t_\omega) - \frac{\alpha^3 \mu_1}{32}\sin(3 t_\omega)
    \\+ \left[
        \alpha \mu_1 \omega_{100} t_\omega
        + \alpha \mu_2 t_\omega \left(\omega_{010} - \frac{3}{8} \alpha^2\right)
        + \alpha \mu_3 \omega_{001} t_\omega
        - \frac{\alpha \mu_1}{2} \left(1 - \frac{7}{16} \alpha^2\right)
    \right] \sin t_\omega.
\end{multline}
By choosing the values of parameters
$$
    \alpha = 2,\quad
    \omega_{100} = 0,\quad
    \omega_{010} = \frac{3}{8} \alpha^2 = \frac{3}{2},\quad
    \omega_{001} = 0,
$$
we then ensure that the so-called \textit{secular} terms proportional to $t_\omega$, which break periodicity of the solution and diverge as $t_\omega\to\infty$, vanish.

With all the above choices we obtain the approximation of the uninodal limit-cycle solution
\begin{multline}
    z_0(t) = \sqrt{-\frac{c_0}{c_2}} q(\omega \sqrt{c}_1 t) \approx \sqrt{-\frac{c_0}{c_2}} \left\{
        \mu_3 + 
        \left(2 - \frac{\mu_2}{4} - \mu_3\right) \cos\left[\sqrt{c_1} t\left(1 + \frac{3}{2} \mu_2\right)\right]
        + \frac{\mu_2}{4} \cos\left[3 \sqrt{c_1} t\left(1 + \frac{3}{2} \mu_2\right)\right]
    \right.\\\left.
        + \frac{3\mu_1}{4} \sin\left[\sqrt{c_1} t\left(1 + \frac{3}{2} \mu_2\right)\right]
        - \frac{\mu_1}{4} \sin\left[3 \sqrt{c_1} t\left(1 + \frac{3}{2} \mu_2\right)\right]
    \right\}.
\end{multline}

%%%
\subsection*{Linear response function}
By using the approximate solution $q(t)$ found above, we now seek the first-order Volterra kernel, also in an approximate form. To this end we first replace the independent variable $t_\omega = \omega s$ with $s$ in Eqs.~[\ref{eq:uni}], and then add a perturbation $C \delta(s)$ with an arbitrary constant $C$ on the right-hand side:
\begin{equation}
     \ddot{q}_C(s) + q_C(s) - \mu_1 \dot{q}_C(s) \left[1 - q_C(s)^2\right] + \mu_2 q_C(s)^3 - \mu_3 = C \delta(s),
\end{equation}
with overdot denoting the differentiation with respect to $s$.
By using the first-order Volterra-series ansatz analogous to Eq.~[\ref{eq:volterra}]
\begin{equation}\label{eq:qvolterra}
    q_C(s) \simeq q(s) + C \int_0^s ds'\, g_q(s - s') \delta(s') = q(s) + C g_q(s),
\end{equation}
with the previously obtained form of $q(s)$ and the linear-response function $g_q(s)$, we further get
\begin{equation}
    C \left[
        \ddot{g}_q - \mu_1 \left(1 - q^2\right) \dot{g}_q
            + \left(1 + 2 \mu_1 q \dot{q} + 3 \mu_2 q^2\right) g_q
    \right] + O(C^2) = C \delta(s),
\end{equation}
which represent a linear equation of Floquet type---with time-dependent coefficients---for $g_q(t)$. As in Ref.~\cite{Belousov2020}, we are going to approximate these coefficients by time averaging
\begin{align}
    &\chi_0 = \mu_1\int_0^{2\pi/\omega} \frac{\omega ds}{2\pi} \left[1 - q(s)^2\right] = \mu_1 \left(1 - 2 \mu_3 + \frac{3}{2} \mu_3^2\right),\\
    &\chi_1 = \int_0^{2\pi/\omega} \frac{\omega ds}{2\pi}
        \left[1 + 2 \mu_1 q(s) \dot{q}(s) + 3 \mu_2 q(s)^2\right] = 1 + 6\mu_2 \left(1 - \mu_3 + \frac{3}{4} \mu_3^2\right),
\end{align}
over the period $2\pi / \omega$. The equation, from which we can find $g_z(t) = \sqrt{-c_0/c_2} g_q(\sqrt{c_1} t)$ (cf. Eq.~[\ref{eq:volterra}] and [\ref{eq:qvolterra}]), then become
\begin{equation}\label{eq:gzbar}
    \sqrt{-\frac{c_0 c_1}{c_2}} \delta(t) = \ddot{g}_z + \sqrt{c_1} \chi_0 \dot{g}_z + c_1 \chi_1 g_z = \ddot{g}_z - c_0 \left(1 - 2 \mu_3 + \frac{3}{2} \mu_3^2\right) \dot{g}_z - 6 \frac{c_0 c_3}{c_2} \left(1 - \mu_3 + \frac{3}{4} \mu_3^2\right).
\end{equation}
Note that as $\mu_1\to0$ and $\mu_2\to0$ the above equation tends to that of a forced undamped harmonic oscillator $g_h = \lim_{\mu_1\to0,\mu_2\to0} g_z$:
\begin{equation}
    \sqrt{-\frac{c_0 c_1}{c_2}} \delta(t) = \ddot{g}_h + c_1 g_h.
\end{equation}

%%% Each figure should be on its own page

\begin{figure}[p!]
\centering
\includegraphics[width=1\columnwidth]{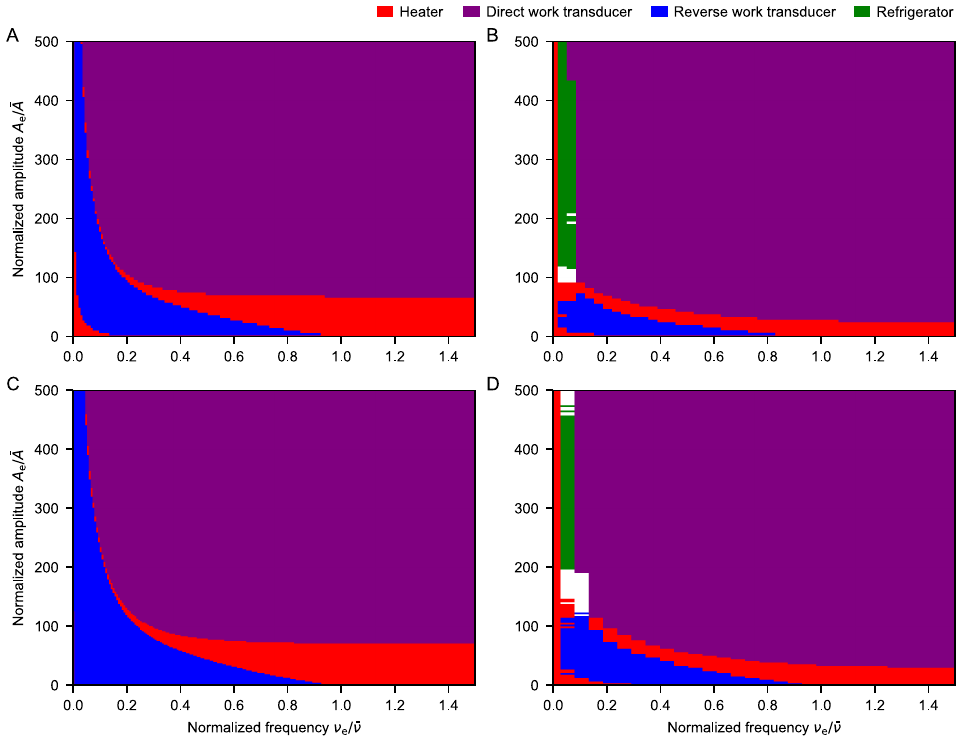}
\caption{Diagram of thermodynamic regimes for hair bundles described by the parameter sets \#1 (A), \#2 (B), \#3 (C), and \#4 (D): direct (purple) and reverse (blue) work-to-work transduction, heater (red), and refrigerator (green). White-colored areas correspond to noisy regimes, in which fluctuations of energy flows are too large to assign one of the listed regimes with confidence. The amplitudes of the external signals are normalized by $\bar{A} = \sqrt{2 \kBT \gamma \bar{\nu}}$, in which the natural frequency $\bar{\nu}$ of the respective parameter set yields $\bar{A}_1 = \SI{0.12 \pm 0.01}{pN}$, $\bar{A}_2 = \SI{0.111 \pm 0.001}{pN}$, $\bar{A}_3 = \SI{0.240 \pm 0.004}{pN}$, and $\bar{A}_4 = \SI{0.171 \pm 0.001}{pN}$ with the natural frequencies $\bar{\nu}_1 = \SI{13.077 \pm 0.001}{Hz}$, $\bar{\nu}_2 = \SI{2.9641 \pm 0.0004}{Hz}$, $\bar{\nu}_3 = \SI{14.6667 \pm 0.0005}{Hz}$, and $\bar{\nu}_4 = \SI{1.8887\pm 0.0002}{Hz}$ of the unperturbed oscillations described by the respective parameter sets. Panel A and B are also shown in Figure 3 of the maintext.}
\label{fig:map}
\end{figure}

\begin{figure}[p!]
\centering
\includegraphics[width=1\columnwidth]{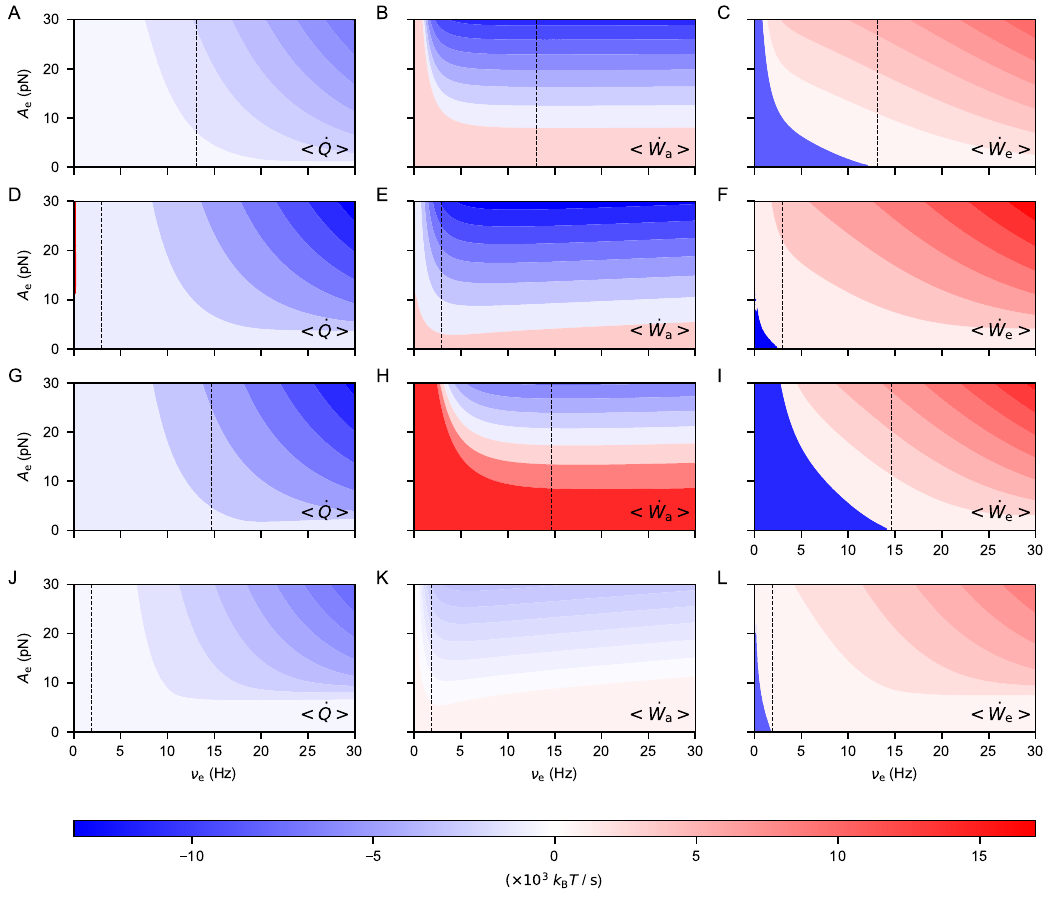}
\caption{Average dissipation rate $\langle \dot{Q} \rangle$ (A,D,G,J), active power $\langle \dot{W}_{\rm a} \rangle$ (B,E,H,K), and external power $\langle \dot{W}_{\rm e} \rangle$ (C,F,I,L) as functions of frequency ($\nu_{\rm e}$) and amplitude ($A_{\rm e}$) of a sinusoidal signal for parameter sets \#1 (A,B,C), \#2 (D,E,F), \#3 (G,H,I), and \#4 (J,K,L). The vertical black dashed line indicates the hair cell's natural frequency.}
\label{fig:flows}
\end{figure}

\begin{figure}[p!]
\centering
\includegraphics[width=1\columnwidth]{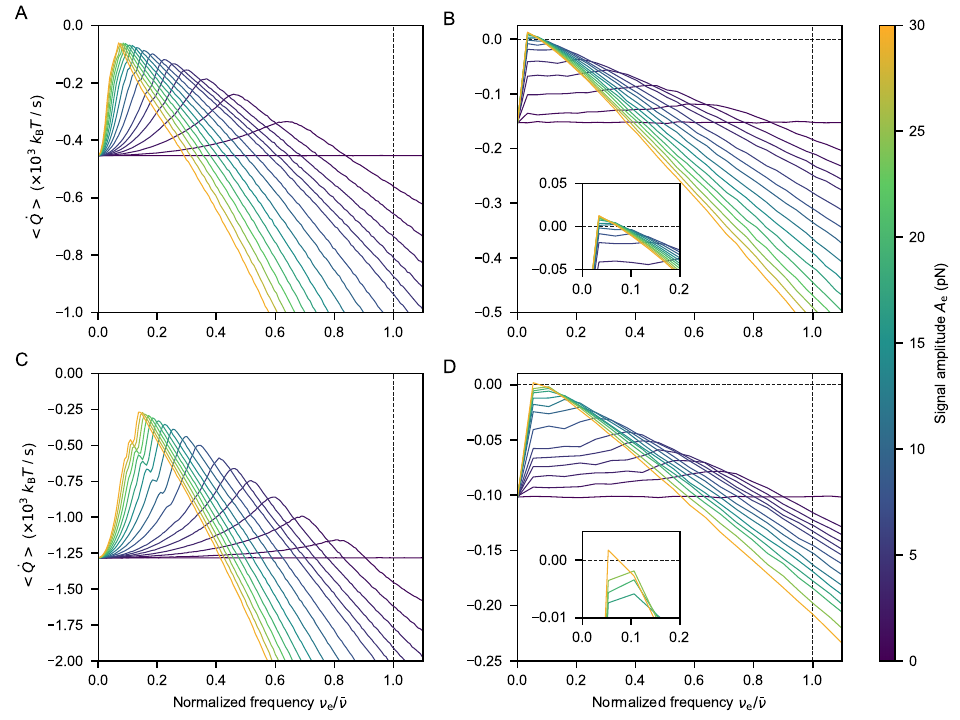}
\caption{Dependence of heat dissipation on the external-signal frequency $\nu_{\rm_e}$ normalized by the natural frequency $\bar{\nu}$ of hair-bundle oscillations (vertical dashed line) for various amplitudes of the stimulus $A_{\rm e}$. Parameter sets: \#1 (A),  \#2 (B), \#3 (C), and \#4 (D).}
\label{fig:q}
\end{figure}

\begin{figure}[p!]
\centering
\includegraphics[width=1\columnwidth]{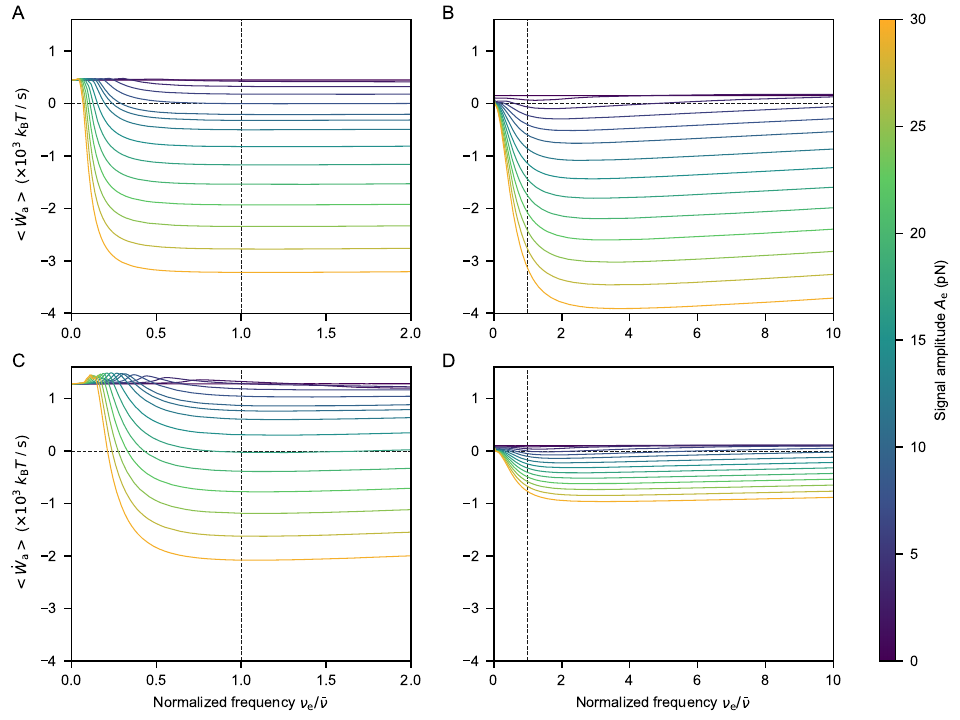}
\caption{Power applied by the hair cell's internal active process as a function of the external signal's frequency $\nu_{\rm e}$ normalized by the cell's natural frequency $\bar\nu$ (vertical dashed line) for various amplitudes of the stimulus $A_{\rm e}$. Parameter sets: \#1 (A),  \#2 (B), \#3 (C), and \#4 (D).}
\label{fig:wa}
\end{figure}

\begin{figure}[p!]
\centering
\includegraphics[width=1\columnwidth]{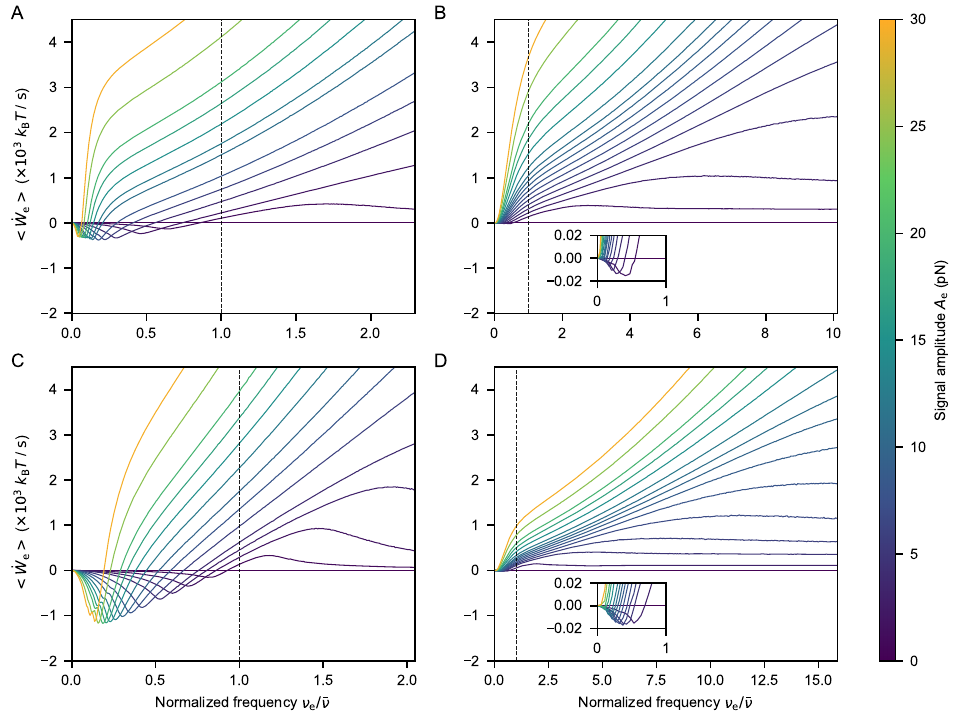}
\caption{Power supplied by the external stimulus as a function of the signal's frequency $\nu_{\rm e}$ normalized by the cell's natural frequency $\bar\nu$ (vertical dashed line) for various amplitudes of the stimulus $A_{\rm e}$. Parameter sets: \#1 (A),  \#2 (B), \#3 (C), and \#4 (D).}
\label{fig:we}
\end{figure}

\begin{figure}[p!]
\centering
\includegraphics[width=1\columnwidth]{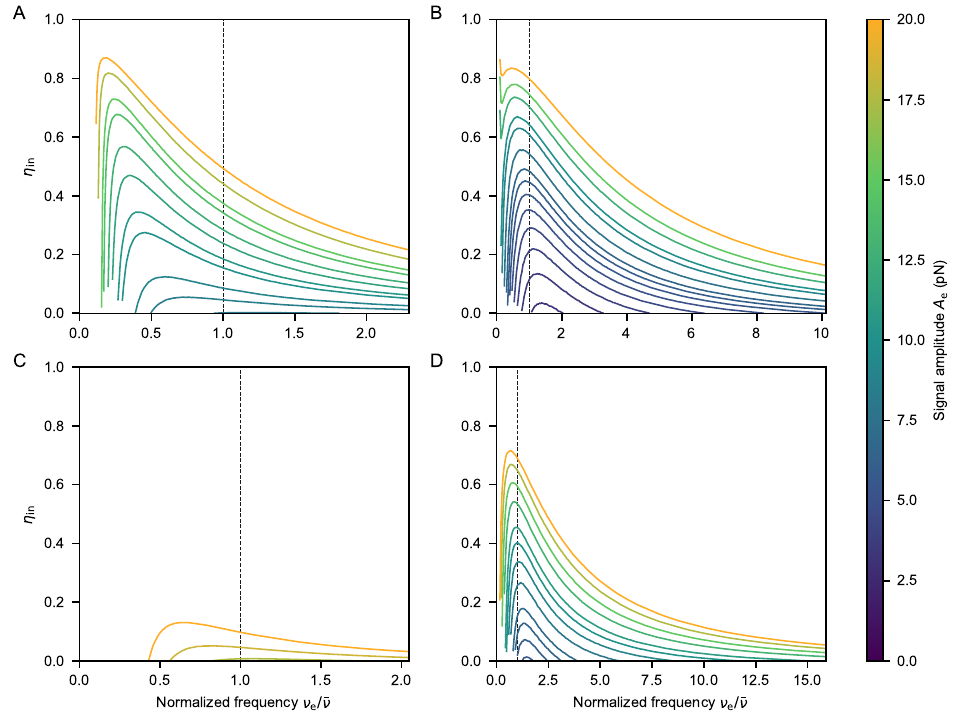}
\caption{Efficiency of the direct work transduction $\eta_{\rm in}$ as a function of the signal's frequency $\nu_{\rm e}$ normalized by the cell's natural frequency $\bar\nu$ (vertical dashed line) for various amplitudes of the stimulus $A_{\rm e}$. Parameter sets: \#1 (A),  \#2 (B), \#3 (C), and \#4 (D). Note that part of this figure, panel A and B, are shown in Figure 5 in the maintext.}
\label{fig:in}
\end{figure}

\begin{figure}[p!]
\centering
\includegraphics[width=1\columnwidth]{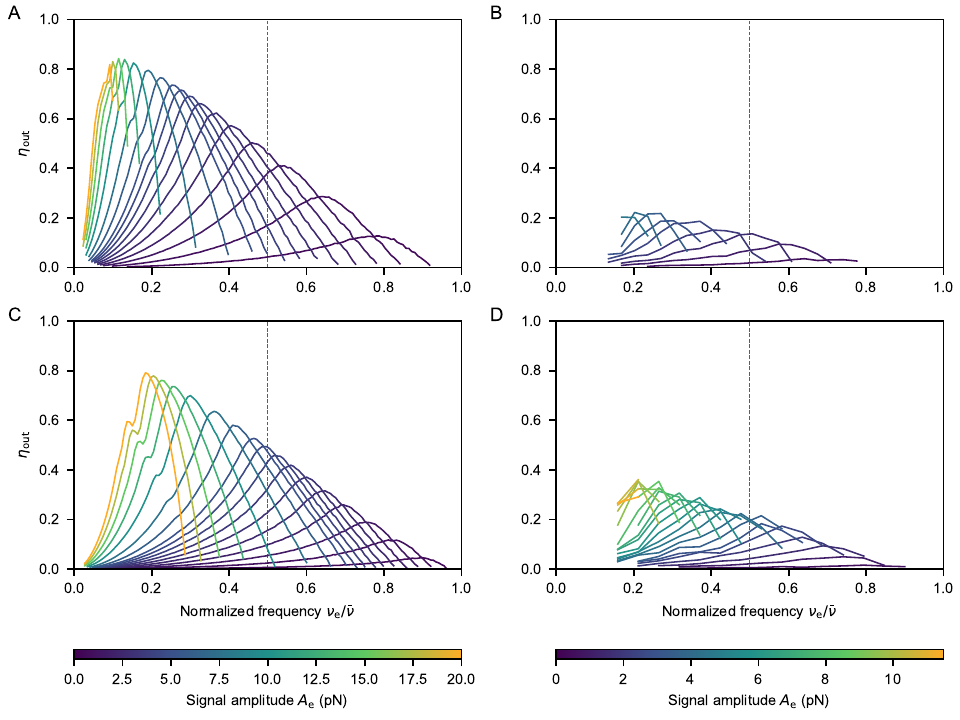}
\caption{Efficiency of reverse work transduction $\eta_{\rm out}$ as a function of the signal's frequency $\nu_{\rm e}$ normalized by the cell's natural frequency $\bar\nu$ for various amplitudes of the stimulus $A_{\rm e}$. Parameter sets: \#1 (A),  \#2 (B), \#3 (C), and \#4 (D). Vertical red dashed line indicates half of the natural frequency. Note that part of this figure, panel A and B, are shown in Figure 5 in the maintext.}
\label{fig:out}
\end{figure}

\begin{figure}[p!]
\centering
\includegraphics[width=1\columnwidth]{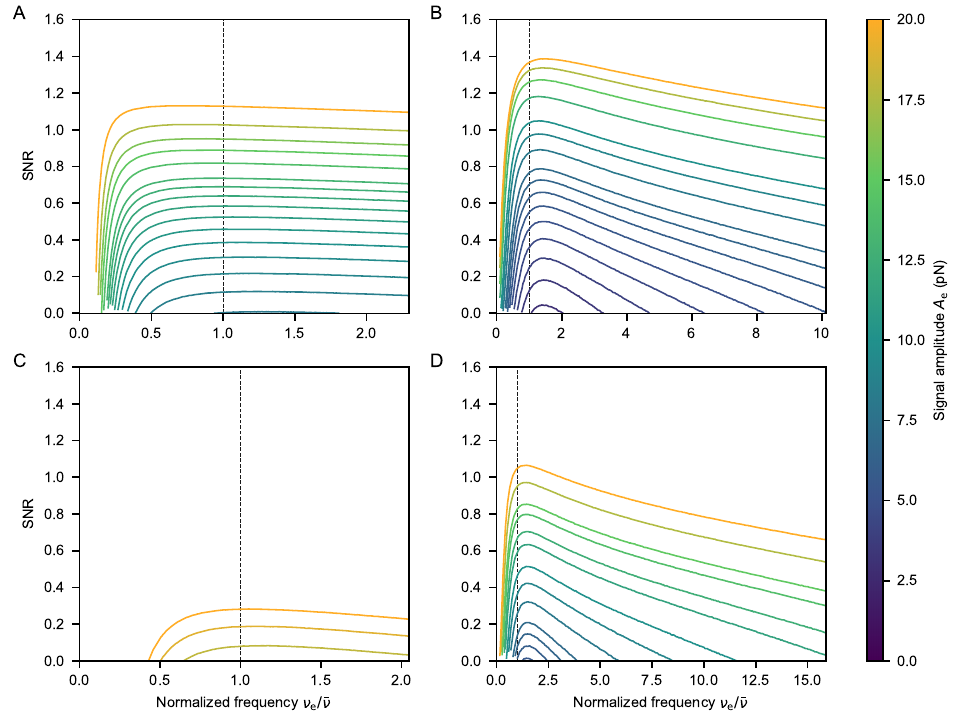}
\caption{Signal-to-noise-ratio (SNR) of the direct work transduction as a function of the signal's frequency $\nu_{\rm e}$ normalized by the cell's natural frequency $\bar\nu$ (vertical dashed line) for various amplitudes of the stimulus $A_{\rm e}$. Parameter sets: \#1 (A),  \#2 (B), \#3 (C), and \#4 (D). Note that part of this figure, panel A and B, are shown in Figure 5 in the maintext.}
\label{fig:snr}
\end{figure}

\begin{figure}[p!]
\centering
\includegraphics[width=1\columnwidth]{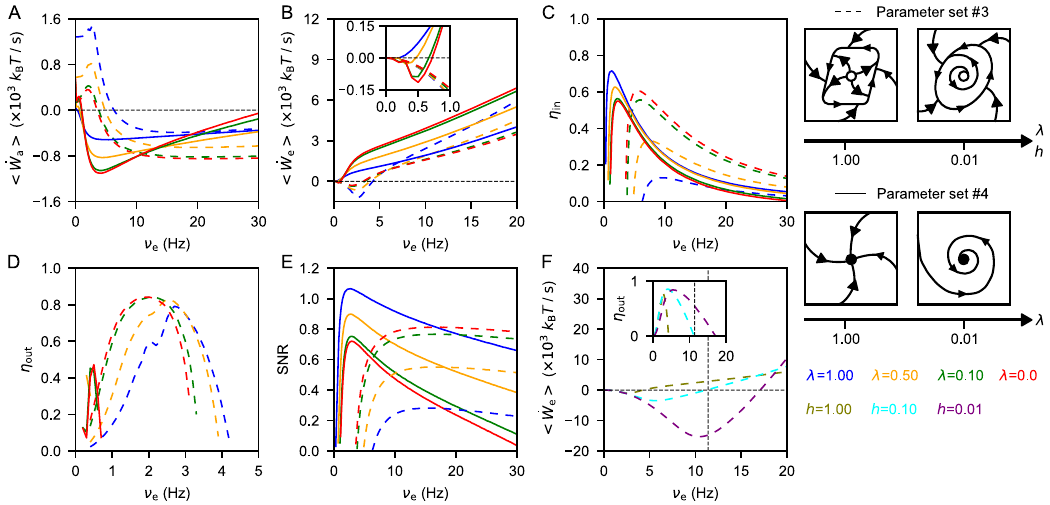}
\caption{Work transduction in different dynamical regimes: parameter sets \#3 (dashed curves, uninodal relaxation oscillations, scaled by $\lambda_3$ or $h$) and \#4 (solid curves, monostabile regime, scaled by $\lambda$) for the external signals of the same amplitude $A_{\rm e} = \SI{20}{pN}$ (A, B) and various frequencies $\nu_{\rm e}$.
A: While approaching the Hopf bifurcation ($\lambda \to 0$), curves describing the active power shift down in both cases, \#3 and \#4, implying more power intake in the direct work transduction and less power expanded in the reverse regime.
B: The curves of the external power become steeper in the proximity of the Hopf bifurcation of the parameter set \#4, indicating that more energy is consumed, but also more is transduced by the reverse work-to-work machine. The opposite trend is observed in the case \#3---the curves flatten.
C: Whereas the power intake of the monostable regime (\#4) increases closer to the Hopf bifurcation, the efficiency of work transduction decreases. The uninodal regime (\#3) transduces much less power, but more efficiently in the proximity of the Hopf bifurcation.
D: At a given amplitude of the external signal, the Hopf bifurcation enables reverse work transduction in the parameter set \#4, and improves its efficiency in the set \#3.
E: The signal-to-noise ratio is maximal in the monostable regime, far from the Hopf bifurcation as $\lambda \to 1$.
F: The low-friction and weak-nonlinearity limit of the Van der Pol -- Duffing Eq.~[\ref{eq:hvdf}] ($h\to0$, case \#3) resembles an underdamped harmonic oscillator. The reverse work transduction is maximal close to the half of the resonance frequency $\nu_0 = \sqrt{c_1} / (2\pi)$. The sketches on the right illustrate how the phase portraits of the system change with the scaling parameters $\lambda$ or $h$, whose values are color-coded as shown in the legend. Effects of these parameters are reported in Tables~\ref{tab:scaling3} and \ref{tab:scaling4}.}
\label{fig:last}
\end{figure}

\clearpage
\begin{table}[p!]\centering
\caption{\label{tab:scaling1}Values of parameter set \#1 scaled by $\lambda$ and $h$ (Fig.~\ref{fig:last})}
\begin{tabular}{ccccccc} 
    \toprule
    Parameter (unit) & $\lambda$ or $h = 1$ & $\lambda = 0.5$ & $\lambda = 0.1$ & $\lambda = 0.01$ & $h = 0.1$ & $h = 0.01$ \\
    \midrule
    $k \, (\si{s^{-1}})$ & $1198.2$ & $1198.2$ & $1198.2$ & $1198.2$ & $1198.2$ & $1198.2$ \\
    $a \, (\si{s^{-1}})$ & $2399$ & $1814.28$ & $1346.50$ & $1241.25$ & $1337.09$ & $1230.9$ \\
    $c \, (\si{{nm}^{-2}\cdot s^{-1}})$ & $1.68$ & $1.68$  & $1.68$ & $1.68$ & $0.168$ & $0.0168$ \\
    $b \, (\si{s^{-1}})$ & $20.9$ & $20.85$ & $20.81$ & $20.80$ & $20.81$ & $20.80$ \\
    $e \, (\si{s^{-1}})$ & $21.0$ & $20.95$ & $20.91$ & $20.90$ & $20.91$ & $20.90$ \\
    $c_y \, (\si{nm \cdot s^{-1}})$ & $28$ & $28$ & $28$ & $28$ & $28$ & $28$ \\
    $c_x \, (\si{nm \cdot s^{-1}})$ & $-23360$  & $-23360$ & $-23360$ & $-23360$ & $-23360$ & $-23360$ \\
    $\gamma \, (\si{\pico N\cdot s\cdot {nm}^{-1}})$ & $0.000127$ & $0.000127$ & $0.000127$ & $0.000127$ & $0.000127$ & $0.000127$ \\
    \bottomrule
\end{tabular}
\end{table}

\begin{table}[p!]\centering
\caption{\label{tab:scaling2}Values of parameter set \#2 scaled by $\lambda$ and $h$ (Fig.~\ref{fig:last})}
\begin{tabular}{ccccc} 
    \toprule
    Parameter (unit) & $\lambda = 1$ & $\lambda = 0.5$ & $\lambda = 0.1$ & $\lambda = 0.01$ \\
    \midrule
    $k \, (\si{s^{-1}})$ & $124.4$ & $124.4$ & $124.4$ & $124.4$ \\
    $a \, (\si{s^{-1}})$ & $619$ & $796.10$ & $937.77$ & $969.65$  \\
    $c \, (\si{{nm}^{-2}\cdot s^{-1}})$ & $0.73$ & $0.73$  & $0.73$ & $0.73$ \\
    $b \, (\si{s^{-1}})$ & $22.59$ & $35.24$ & $45.37$ & $47.65$ \\
    $e \, (\si{s^{-1}})$ & $31.48$ & $44.14$ & $54.26$ & $56.54$\\
    $c_y \, (\si{nm \cdot s^{-1}})$ & $225.4$ & $225.4$ & $225.4$ & $225.4$ \\
    $c_x \, (\si{nm \cdot s^{-1}})$ & $1528$  & $1528$ & $1528$ & $1528$ \\
    $\gamma \, (\si{\pico N\cdot s\cdot {nm}^{-1}})$ & $0.000511$ & $0.000511$ & $0.000511$ & $0.000511$ \\
    \bottomrule
\end{tabular}
\end{table}

\begin{table}[p!]\centering
\caption{\label{tab:scaling3}Values of parameter set \#3 scaled by $\lambda$ and $h$ (Fig.~\ref{fig:last}).}
\begin{tabular}{ccccccc} 
    \toprule
    Parameter (unit) & $\lambda$ or $h = 1$ & $\lambda = 0.5$ & $\lambda = 0.1$ & $\lambda = 0.01$ & $h = 0.1$ & $h = 0.01$ \\
    \midrule
    $k \, (\si{s^{-1}})$ & $677$ & $677$ & $677$ & $677$ & $677$ & $677$ \\
    $a \, (\si{s^{-1}})$ & $1300$ & $1003.98$ & $767.17$ & $713.89$ & $766.70$ & $713.37$ \\
    $c \, (\si{{nm}^{-2}\cdot s^{-1}})$ & $0.914$ & $0.914$  & $0.914$ & $0.914$ & $0.0914$ & $0.00914$ \\
    $b \, (\si{s^{-1}})$ & $30.42$ & $30.34$ & $30.28$ & $30.26$ & $30.28$ & $30.26$ \\
    $e \, (\si{s^{-1}})$ & $30.60$ & $30.52$ & $30.46$ & $30.44$ & $30.46$ & $30.44$ \\
    $c_y \, (\si{nm \cdot s^{-1}})$ & $13.1$ & $13.1$ & $13.1$ & $13.1$ & $13.1$ & $13.1$ \\
    $c_x \, (\si{nm \cdot s^{-1}})$ & $-564$  & $-564$ & $-564$ & $-564$ & $-564$ & $-564$ \\
    $\gamma \, (\si{\pico N\cdot s\cdot {nm}^{-1}})$ & $0.00048$ & $0.00048$ & $0.00048$ & $0.00048$ & $0.00048$ & $0.00048$ \\
    \bottomrule
\end{tabular}
\end{table}

\begin{table}[p!]\centering
\caption{\label{tab:scaling4}Values of parameter set \#4 scaled by $\lambda$ and $h$ (Fig.~\ref{fig:last}).}
\begin{tabular}{ccccc} 
    \toprule
    Parameter (unit) & $\lambda = 1$ & $\lambda = 0.5$ & $\lambda = 0.1$ & $\lambda = 0.01$ \\
    \midrule
    $k \, (\si{s^{-1}})$ & $136.9$ & $136.9$ & $136.9$ & $136.9$ \\
    $a \, (\si{s^{-1}})$ & $759$ & $1001.94$ & $1196.29$ & $1240.02$  \\
    $c \, (\si{{nm}^{-2}\cdot s^{-1}})$ & $3.69$ & $3.69$  & $3.69$ & $3.69$ \\
    $b \, (\si{s^{-1}})$ & $15.12$ & $24.56$ & $32.11$ & $33.81$ \\
    $e \, (\si{s^{-1}})$ & $20.44$ & $29.88$ & $37.43$ & $39.13$\\
    $c_y \, (\si{nm \cdot s^{-1}})$ & $57.3$ & $57.3$ & $57.3$ & $57.3$ \\
    $c_x \, (\si{nm \cdot s^{-1}})$ & $264$  & $264$ & $264$ & $264$ \\
    $\gamma \, (\si{\pico N\cdot s\cdot {nm}^{-1}})$ & $0.001897$ & $0.001897$ & $0.001897$ & $0.001897$ \\
    \bottomrule
\end{tabular}
\end{table}

\end{document}